\PassOptionsToPackage{hypertexnames=false}{hyperref}
\documentclass[pdflatex,sn-mathphys-num]{sn-jnl}
\usepackage[T1]{fontenc}
\usepackage{graphicx}%
\usepackage{multirow}%
\usepackage{amsmath,amssymb,amsfonts}%
\usepackage{amsthm}%
\usepackage{xcolor}%
\usepackage{textcomp}%
\usepackage{manyfoot}%
\usepackage{booktabs}%
\usepackage{tabularx}%
\usepackage{array}%
\usepackage{xurl}
\usepackage{float}       
\usepackage{tcolorbox}   
\tcbuselibrary{skins}    
\usepackage{enumitem}    

\theoremstyle{thmstyleone}%

\theoremstyle{thmstyletwo}%

\theoremstyle{thmstylethree}%

\begin{document}

\title[LLM-Agent Social Simulations]{Towards Operational Validation of LLM-Agent Social Simulations: A Replicated Study of a Reddit-like Technology Forum}


\author*[1]{\fnm{Aleksandar} \sur{Toma\v{s}evi\'{c}}}\email{atomasevic@ipb.ac.rs}

\author[1]{\fnm{Darja} \sur{Cvetkovi\'{c}}}

\author[2]{\fnm{Sara} \sur{Major}}

\author[3]{\fnm{Slobodan} \sur{Maleti\'{c}}}

\author[3]{\fnm{Miroslav} \sur{An\dj{}elkovi\'{c}}}

\author[1]{\fnm{Ana} \sur{Vrani\'{c}}}

\author[1]{\fnm{Boris} \sur{Stupovski}}

\author[1]{\fnm{Du\v{s}an} \sur{Vudragovi\'{c}}}

\author[1]{\fnm{Aleksandar} \sur{Bogojevi\'{c}}}

\author[1]{\fnm{Marija} \spfx{Mitrovi\'{c}} \sur{Dankulov}}

\affil*[1]{\orgname{Institute of Physics Belgrade}, \orgdiv{University of Belgrade}, \orgaddress{\city{Belgrade}, \country{Serbia}}}

\affil[2]{\orgname{Faculty of Philosophy}, \orgdiv{University of Novi Sad}, \orgaddress{\city{Novi Sad}, \country{Serbia}}}

\affil[3]{\orgname{Vin\v{c}a Institute of Nuclear Sciences}, \orgdiv{University of Belgrade}, \orgaddress{\city{Belgrade}, \country{Serbia}}}


\abstract{\label{pg:abstract}Validation of LLM-agent social simulations remains underdeveloped, with most studies relying on subjective assessments or single runs. We address this gap by running 30 independent 30-day simulations of a technology forum modeled on Voat’s v/technology, using stateless Dolphin Mistral 24B agents on the Y Social platform, and evaluating operational validity across five dimensions: activity patterns, network structure, toxicity, topical coverage, and stylistic convergence. Against 30 matched, non-overlapping 30-day Voat comparison windows, results show overlapping 99\% confidence intervals for unique users, root posts, and daily active users, while comments, average thread length, and mean toxicity remain higher in simulation. Both simulated and empirical networks exhibit core--periphery structure, though simulated cores are larger and more diffuse and repeated interactions are less frequent. Topic alignment is near-complete, but toxicity is misallocated across content layers: simulated root posts are substantially more toxic than real submissions, while simulated comments are less toxic than Voat comments. These findings demonstrate that LLM agents in platform-faithful environments can reproduce familiar online regularities, while systematic divergences---particularly those linked to stateless agent design and content-layer calibration---point to concrete directions for future improvement.}


\maketitle

\section{Introduction}
\label{sec:introduction}

Online social media is a primary arena for public discourse and collective sense-making, yet it often generates polarization, echo chambers, harassment, and norm violations that co-evolve with moderation and platform design \cite{cinelli2021echo,avalle2024persis,pandita2024roots,neuman2024affect}.
Studying such dynamics \textit{in vivo} is increasingly difficult in the post-API era due to ethical issues, limited access, and reproducibility constraints \cite{mimizuka2025postapi}. Complementary to traditional agent-based models, recent LLM-based simulations embed language-model agents in platform-faithful environments to test whether recognizable online regularities emerge from culturally informed, norm-conditioned interactions \cite{anthis2025llm-a,vezhnevets2023genera,larooij2025do}. However, generating realistic contentious and toxic discourse remains challenging: standard aligned models tend toward politeness and consensus, poorly suited to simulate the disagreements and norm violations that drive real online dynamics. Moreover, generative agent-based modeling faces a central methodological challenge: rigorous validation frameworks remain underdeveloped \cite{larooij2025do,li2024compre}. Recent reviews find that most studies rely on subjective ``believability'' assessments rather than quantitative comparison with empirical data, and nearly all report results from single simulation runs, akin to drawing conclusions from a single case study \cite{larooij2025do}. Newly emerging proposals thus caution against treating LLM outputs as substitutes for human participants without explicit empirical checks \cite{adornetto2025genera,lin2025fallacies}. Even when multi-user discussions appear realistic, conversational realism does not guarantee that platform-level structures and distributions match the target system \cite{bouleimen_collective_turing,mayor2025spoken,ngcarley_llm_bots}. Ground truth for validating that agents act as the persons they simulate is rarely accessible, and the black-box nature of LLMs intensifies interpretability problems \cite{adornetto2025genera}.

As online platforms begin to mix people and artificial users in the future, human-machine interactions will become an increasingly important subject of social-science research \cite{tsvetkova2024new}. Understanding what AI agents alone can produce in a social media environment provides a critical baseline for later human-machine studies, where even suspected machine presence may change how people act. The present work presents a first step toward this goal, by studying machine-machine interaction in a Reddit-like forum. We treat LLM agents as carriers of cultural knowledge and conversational norms \cite{brinkmann2023machin-a}, and ask whether they can share, remix, and react to content in ways that yield familiar social-media patterns, without agents being given a specific aim or over-control of their behavior. 

We view LLM agents as cultural technologies rather than payoff‑maximizing decision makers. Agents do not optimize utilities for posting, replies, or visibility. Instead, models trained on large conversational corpora encode social norms, roles, and routines. In practice, agents respond by what seems \textit{appropriate} in context—given a topic, a thread position, and a persona—rather than by trying to achieve a specific goal through reward maximization. This lets us test whether compressed cultural knowledge alone can generate familiar online patterns when placed inside a realistic platform setting \cite{brinkmann2023machin-a,farrell2025large,vezhnevets2023genera}.

Concretely, we simulate Voat’s v/technology forum on the Y Social platform using persona-based LLM agents and a fixed catalog of technology URLs drawn from the original forum. To preserve disagreement and the possibility of toxic speech, we use the uncensored Dolphin Mistral 24B model. Voat offers a useful baseline because its Reddit-like architecture and complete lifecycle data enable platform-faithful calibration, while v/technology itself is comparatively mainstream in topic content on the platform, though not necessarily in ideological composition.

\begin{figure}[ht]
\centering
\includegraphics[width=\textwidth]{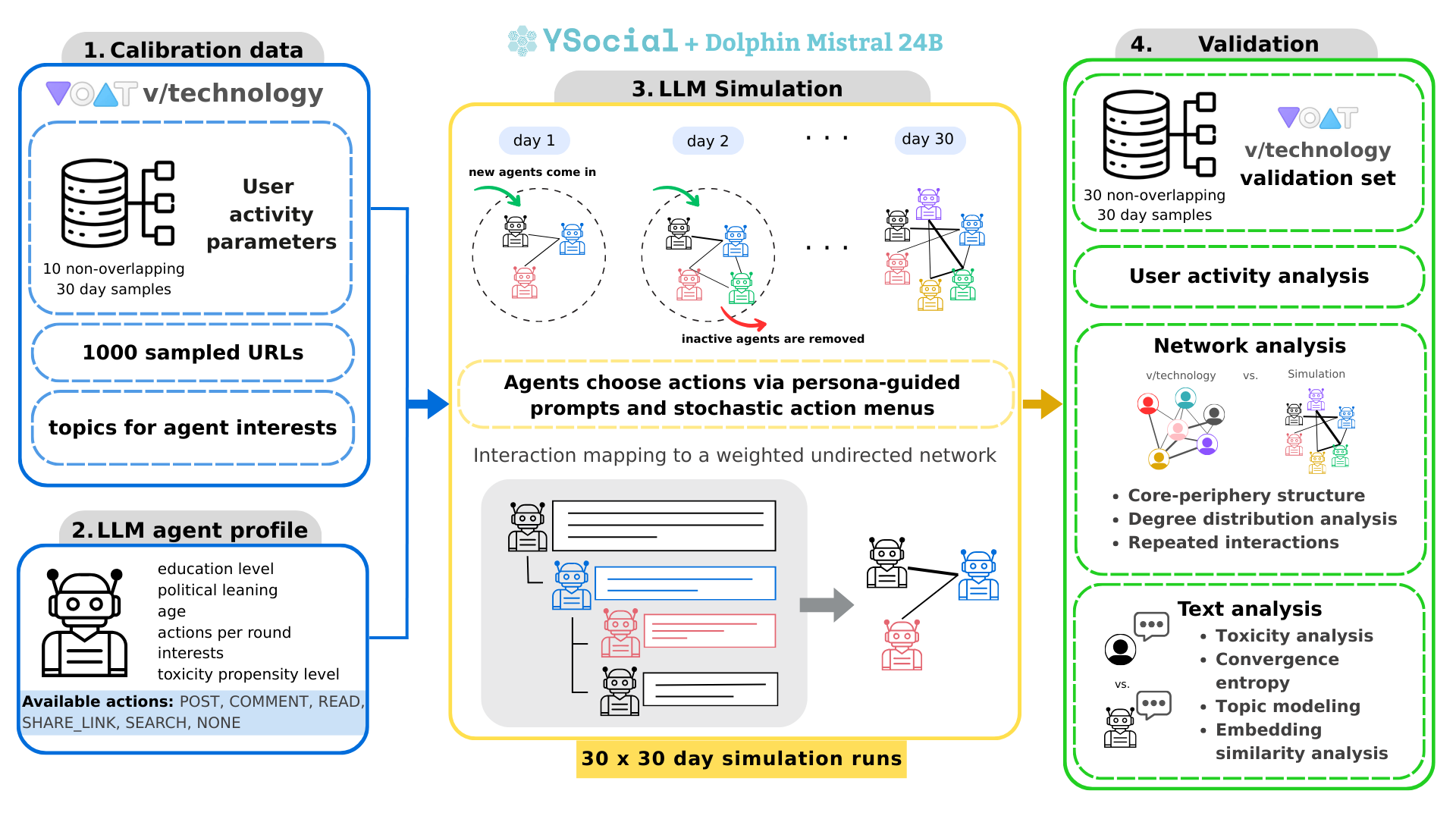}
\caption{\textbf{Visual abstract of the simulation and validation workflow.}
The study proceeds through four stages.
\textbf{(1) Calibration data:} we extract user activity parameters from 10 non-overlapping 30-day samples of Voat's v/technology subverse (MADOC dataset).
\textbf{(2) LLM agent profile:} each agent is assigned a concise persona comprising education level, political leaning, age, per-round action budget, topical interests, and a toxicity propensity level.
\textbf{(3) LLM simulation:} the Y Social platform runs 30 independent 30-day simulations using Dolphin Mistral 24B; agents enter and exit the population daily according to calibrated growth and churn rates, choose actions via persona-guided prompts and stochastic menus, and their interactions are mapped to a weighted undirected network.
\textbf{(4) Result validation:} simulation outputs are compared against 30 matched, non-overlapping 30-day Voat windows across five dimensions: user activity patterns, network structure (core-periphery, degree distribution, repeated interactions), and text analysis (toxicity, convergence entropy, topic modeling, embedding similarity).}
\label{fig:voat_simulation}
\end{figure}

We calibrate population size, growth, posting and reply likelihoods, and per-round action budgets to Voat’s v/technology and run 30-day simulations. We adopt a complex‑systems view \cite{larooij2025do,anthis2025llm-a}: can LLM agents, carrying compressed cultural knowledge and operating under realistic platform constraints, reproduce the emergent patterns characteristic of online communities?


We address this question through three validation dimensions, which also structure our contributions:

\begin{enumerate}
\item \textbf{Activity patterns:} Do simulated posting rates, thread dynamics, and user turnover match empirical distributions? We provide a platform‑faithful baseline calibrated to Voat's v/technology with replicated validation across 30 runs.

\item \textbf{Network structure:} Does the interaction network exhibit comparable structural properties? We analyze core–periphery organization, degree distributions, and repeated interactions against 30 Voat comparison windows.

\item \textbf{Textual convergence:} Do agent outputs converge toward reference community characteristics? We evaluate toxicity distributions, topic alignment via embedding similarity, and linguistic convergence entropy.
\end{enumerate}

Together, these provide an operational-validation framework and a reproducible approach to LLM-based community simulation. Figure~\ref{fig:voat_simulation} provides a visual overview of the complete workflow.

The paper proceeds as follows: we first discuss our view of LLMs as social agents, then outline the simulation approach and design, detail evaluation methods, present results on activity, network structure, toxicity, topics/embeddings, and stylistic convergence, and discuss implications and limitations.

\section{LLMs as social agents}
\label{sec:llms-social}

Recent work argues that large language models are best understood as cultural and social technologies: systems that reorganize and render usable the information produced by people, just as print, markets, and bureaucracies have historically done \cite{farrell2025large}. From this perspective, the key question is not whether models can be autonomous minds, but how they moderate access to, and coordination of, cultural repertoires and human knowledge. Much like earlier technologies, LLMs reduce complexity and function as mediators between individual actors and a broader repository of human knowledge and culture, simultaneously being shaped by existing epistemic and symbolic structures and reciprocally influencing them \cite{brinkmann2023machin-a}. Their ability to compress and disseminate information and cultural scripts, while continually being refined by external input (primarily through reinforcement learning from human feedback) positions them as potentially central to processes of meaning-making in mediated publics. This implies a broad array of potential applications and risks yet to be fully understood \cite{farrell2025large}.

What makes large models a particularly valuable resource for researchers seeking to understand social patterns is their ability to not only store, reproduce, and transform vast amounts of data, but, with careful prompting, to explicate and simulate the various social perspectives implicit in it \cite{park_social_2022, adornetto2025genera}. Consider the action space agents must navigate to produce realistic threads in a Reddit-like environment: a single discussion about operating-system updates might involve quoting prior turns, deploying in-group jargon (kernel, distro, drivers, telemetry), using irony (``just use Arch''), and sustaining disagreement via reply chains, while reactions and feed visibility concentrate attention on the most contested subthreads. Classical ABMs often used parsimonious, deterministic, or rational‑choice rules for tractability, but these are poorly suited for such culture‑laden, role‑dependent interaction \cite{kozlowski2025simula}. LLM simulations replace payoff optimization with linguistic, norm‑guided action selection. This allows them to capture the compression dynamics underlying human social coordination \cite{durrheim2025human}, i.e. the patterned reductions that emerge from the ways human groups process and distill cultural and cognitive complexity. Recent work in generative agent-based modeling has demonstrated the potential for language-mediated simulations \cite{vezhnevets2023genera}, though our approach more specifically targets social media dynamics.

We adopt this norm‑guided approach to LLM simulations, positing that agents' choices are generated via normative role expectations. By “LLM social agent” we mean a language‑model‑driven (artificial) participant in an online forum that produces context‑appropriate posts and replies grounded in shared norms, roles, and routines. Here, the concept of \textit{appropriateness} implies that agent action is grounded in socially agreed-upon rules and understandings about what normal, reasonable, and legitimate behavior is for actors within a given social setting \cite{goodin_logic_2013}. Instead of pursuing utility or optimization of returns, agents must favor normative fit and try to adhere to the prescribed roles and expectations associated with their community. 

Since models are trained on human culture, algorithmic fidelity implies that subpopulation‑conditioned agents can inherit group‑level beliefs and biases. This means they reproduce the social constructions—cultural codes, conventions, and inclinations—already present in the training data, rather than simulating generic human behavior. At the same time, because their “social cognition” is not experiential, they lack the ability to independently reinterpret norms and negotiate social constructs \cite{farrell2025large}, meaning they can only mirror the objectified realities established by humans through socialization and social interaction \cite{berger_social_1991}. By replicating existing cultural narratives, the inherent biases of agents thus closely resemble the real-world propensities of human actors within a given context \cite{argyle2023out}. This allows LLMs to act as carriers of established hermeneutic structures and enables concrete computational implementations of social construction \cite{bao2025langua,vezhnevets2023genera}. 

In Reddit‑like forums where interaction is language‑mediated and governed by explicit rules and tacit norms, we find this framing fits better than homo‑economicus assumptions: agents generate choices by appropriateness and role, leveraging background cultural knowledge to ground identities, routines, and sense‑making \cite{brinkmann2023machin-a,vezhnevets2023genera,anthis2025llm-a}. For instance, under a post about a Windows update and telemetry changes, an open-source/privacy‑oriented persona may argue for Linux alternatives (e.g., Fedora/Debian) and emphasize control and transparency, while a gaming‑oriented persona may defend Windows on the grounds of driver support and DirectX compatibility.

In line with ABM validation practice simulation succeeds \cite{epstein1999generative_social_science,fagiolo2007critical_validation,guerini2017method}, if it reproduces the right distributional patterns for the right reasons, not by matching individuals or text.  Accordingly, we treat LLM agents as tools for studying norm-guided interactions, not as models of human cognition \cite{park2023genera,vezhnevets2023genera}. Building on prior work, first, we design  LLM agents’ interactions as stateless micro-dialogues, without long-term chat memory.



Second, we expose content through a reverse chronological feed based on post popularity (ranked by reaction count with recency tie-breaking) within a 96‑round visibility window, so engaged items are surfaced while decisions remain grounded in recent context. Third, we enforce a small stochastic action menu with heavy‑tailed per‑round budgets to induce unequal participation. Fourth, we prioritize mentions so local conversational dependencies propagate within threads even under stateless prompting. Fifth, we seed discussion from a fixed Voat link catalog to anchor topics culturally without external drift.



\section{Methods}
\label{sec:methods}

We implement the study using Y Social, a client–server digital twin of a Reddit-like forum \cite{rossetti2024y}. The platform consists of three main components: (i) a stateful server that maintains users, content, and recommendation logic; (ii) a simulation engine that advances the simulation clock, schedules agents, and orchestrates interactions; and (iii) a stateless LLM (via Ollama) that generates agent actions and drafts text for posts and comments based on constraints provided by the engine. This architecture enables controlling global parameters such as feed design, initial agent population, simulation duration, activity probabilities, and churn or growth dynamics, while keeping the agent model fixed. At the same time, it logs the full system trajectory for evaluation. Implementation details of Y Social are beyond the scope of this paper; we refer readers to \cite{rossetti2024y} for a detailed description.

\subsection{Calibration Data}
\label{subsec:calibration-data}
To inform the simulation design, we analyze samples from the MADOC dataset, drawn from real-world Voat data, with a focus on the v/technology subverse. This community centers on technology-related topics (e.g., Big Tech, AI, privacy, and gadgets), in contrast to the more political or conspiratorial content associated with other subverses. 

We construct 10 non-overlapping 30-day calibration windows and, for each, compute daily activity, user dynamics, and thread characteristics. Table~\ref{tab:madoc-calib-voat} summarizes the key metrics used to calibrate the simulation (see also Figure~\ref{fig:voat_simulation}, panel 1). Validation in Results uses a separate set of 30 matched, non-overlapping comparison windows (each a single 30-day sample).

\begin{table}[t]
\centering
\caption{MADOC calibration samples summary for Voat v/technology (10 windows of 30 days each). Means and standard deviations are across windows.}
\label{tab:madoc-calib-voat}
\small
\begin{tabular}{@{} l r r r @{}}
\toprule
\textbf{Metric} & \textbf{Mean} & \textbf{SD} & \textbf{Min--Max} \\
\midrule
Users per 30\,days sample & 576.10 & 111.11 & 385--721 \\
Active users per day & 31.52 & 5.96 & 21.50--40.57 \\
New users per day (\%) & 59.44 & 2.29 & 55.69--62.49 \\
Churned users per day (\%) & 75.13 & 1.73 & 71.95--76.80 \\
Comments per post & 1.07 & 0.09 & 0.96--1.19 \\
Posts per 30\,d  & 618.40 & 109.69 & 440--819 \\
Comments per 30\,d & 664.50 & 135.36 & 435--864 \\
Active users on day 1 & 32.60 & 15.05 & 14--66 \\
\bottomrule
\end{tabular}
\end{table}

Taken together, these statistics describe a small and relatively volatile forum: approximately 576 unique users per 30-day window, with around 32 active users per day. Threads are shallow ($\approx$ 1.07 comments per post), and overall activity is low ($\sim$618 posts and $\sim$665 comments per window).  Daily inflow of new users is high ($\sim$60\%), and churn is also high ($\sim$75\%), indicating less persistent engagement. Accordingly, our simulation emulates a smaller, less stable forum with higher turnover and shorter, sparser conversations.

To ensure discussions are oriented toward technological topics, we built a fixed-URL catalog from Voat’s v/technology dataset (MADOC). URLs are cleaned (tracking removed, domains normalized), deduplicated, and sampled to create a pool of 1,000 links (Table \ref{tab:calib-domains}). These are stored locally and used during share-link actions.

\begin{table}[t]
\centering
\caption{Top calibration domains (from calibration URL list). Examples illustrate the kind of content used to seed/calibrate topics.}
\label{tab:calib-domains}
\small
\begin{tabularx}{\textwidth}{@{} l r >{\raggedright\arraybackslash}X @{}}
\toprule
\textbf{Domain} & \textbf{Count} & \textbf{Example (brief description)} \\
\midrule
\texttt{wikipedia.org} & 170 & Commodore International; Van Eck phreaking (encyclopedic background on legacy computing and security) \\
\texttt{github.com} & 62 & AdNauseam FAQ (ad\,tech resistance; extension documentation) \\
\texttt{bitchute.com} & 60 & Platform videos (alternative video hosting; tech/policy adjacent links) \\
\texttt{twitter.com} & 31 & Social link (tweet/profile referenced in discussion) \\
\texttt{hooktube.com} & 26 & Alternative YouTube front\,end (example video link) \\
\texttt{reddit.com} & 21 & r/zeronet (discussion hub for decentralized web) \\
\texttt{breitbart.com} & 17 & Policy/politics piece (tech\,adjacent regulatory context) \\
\texttt{puri.sm} & 14 & Librem 5 smartphone (privacy\,focused hardware product page) \\
\texttt{thepostmillennial.com} & 3 & Opinion/news piece (tech‑adjacent policy/culture coverage) \\
\bottomrule
\end{tabularx}
\end{table}

\subsection{LLM user profiles}
\label{subsec:llm-user-profiles}

In the simulation, the client instantiates a population of agents with sampled personas according to Table ~\ref{tab:agent-init} (see also Figure~\ref{fig:voat_simulation}, panel 2). Each agent is assigned a fixed locale (English, U.S.), an education level (from high school to PhD), an age between 18 and 60, and 2–5 topical interests drawn from a 10-category, technology-focused catalog. Agents are also assigned a toxicity propensity level, {Absolutely No, No, Moderately}, with probabilities (0.80, 0.15, 0.05). Personas generated like this are constructed for face validity rather than demographic representativeness and are not calibrated to Voat’s user population. 

Political leaning is the only benchmark-informed constrained attribute; other persona dimensions are sampled broadly. To reflect the ideological profile of v/technology, agents are sampled from four right-of-center segments listed in Table~\ref{tab:agent-init}, with fixed proportions (0.35, 0.25, 0.25, 0.15). These segments correspond to Pew’s 2021 typology \cite{pewResearchCenter2021beyond} and are intended to approximate the platform’s alt-right–skewed environment \cite{mekacher2022Other-a}. The weights are heuristic and used to induce plausible stance variation, not to estimate Voat’s true population.

Finally, each agent receives a per-round action budget between 1 and 10 actions, sampled from a truncated Zipf distribution (s=2.5). This produces a heavy-tailed activity pattern, most agents are minimally active, while a small minority contribute disproportionately, consistent with participation inequality observed in online communities \cite{panek2017growth,muchnik2013origins}.

\begin{table}[t]
\centering
\caption{Agent population initialization parameters.}
\label{tab:agent-init}
\small
\begin{tabularx}{\textwidth}{@{} l >{\raggedright\arraybackslash}X @{}}
\toprule
\textbf{Attribute} & \textbf{Values / Sampling} \\
\midrule
Locale & English (American) \\
Education level & \{high school, bachelor, master, phd\} \\
Political leaning & Religious-Patriot Conservatives, Pro-Business
Establishment Right, Anti-Elite Populist Right, Socially Moderate
Right \\
Leaning fractions & 0.35, 0.25, 0.25, 0.15 \\
Age & 18--60 \\
Actions per activation (round) & 1--10 \\
Number of interests & 2--5 \\
Interests catalog & Social Media \& Online Platforms; Internet Policy \& Regulation; Artificial Intelligence; Electric Vehicles \& Transportation; Software Development; Clean Energy \& Sustainability; Cybersecurity \& Privacy; Big Tech; Space Technology; Open Source Projects \\
Toxicity propensity level & \{Absolutely No, No, Moderately\}; weights 0.80, 0.15, 0.05 \\
\bottomrule
\end{tabularx}
\end{table}

Each agent is able to perform the following actions:
\begin{enumerate}
\item \textbf{posting} creates a root submission and expands the content inventory; 
\item \textbf{commenting} deepens a thread and alters its shape and lifetime; 
\item \textbf{reading} represents lurking behavior and can shift interests based on consumed content;
\item \textbf{searching} steers engagement toward topical items (e.g., searching for Linux drivers or AI chips) and contributes to topic–interest alignment;
\item \textbf{share-link} propagates existing news posts or ingests fresh articles from the news database, seeding discussion around real‑world content and changing the mix of link vs.\ text submissions.
\end{enumerate}

Interests therefore form a simple feedback loop: they shape exposure and search, are injected into prompts through server-side state, and can themselves shift after engagement.

Agents interact with the platform through short, stateless micro-dialogues between an agent and a handler. The handler, implemented by the simulation engine, prompts the LLM to select actions and generate text. All agent decisions, picking actions from the menu, generating text, and reacting to content, are made by Dolphin Mistral 24B Venice Edition (model details are provided in SI). Importantly, agents have no internal memory; continuity is maintained via server-side state (e.g., thread history and user interests) passed into each prompt. They are also linked to a server-side recommender that provides content across rounds. The feed is non-personalized and ranked by popularity within a fixed visibility window: posts are ordered by total reactions (likes + dislikes), with ties broken by recency. Agents do not see their own posts, and items older than 96 rounds are excluded. When agents open a thread, the visible reply context is capped at \(K=3\) prior items. This uniform cap is a pragmatic context-window control for the smaller model setup rather than a behavioral claim that all users read equally deep. It keeps conversations grounded in the local thread context and the agent's persona. Variable-depth policies tied to agent activity or empirical thread-depth distributions are a natural future extension.

\subsection{LLM simulation}
\label{subsec:llm-simulation}

The simulation runs in discrete daily iterations. At $t=0$, we set the total duration of the simulation and initialize the population with 50 agents generated as described previously. Each day, a subset of agents is activated, and at the end of day $t$, we update the population using the rates from Table~\ref{tab:sim-params}. First, a fraction of agents with the longest inactivity is removed (ties broken randomly). Then, new agents are added, sampled using the same initialization procedure, at a rate proportional to the pre-churn population.

The engagement of active agents is defined by weighted likelihoods, Table \ref{tab:sim-params}. The agent receives a small stochastic menu of actions, and via a prompt, selects one action at a time until its per-round budget is exhausted. Every turn presents exactly three options: \texttt{NONE} and two distinct actions sampled (without replacement) from five options—posting, reading, commenting, searching, and sharing links—according to engagement-likelihood weights. Chosen weights can reproduce the observed comment-to-post ratio, but the mapping from per-turn menus to realized comment volumes is not linear due to reply cascades and thread depth. Passive engagement, such as exposure without content creation, is not directly observable in the calibration data, so the read/search weights must be treated as heuristic. MADOC includes only a static subscriber-count field rather than a time-varying membership series, so it cannot support direct calibration of passive engagement dynamics.

\begin{table}[ht!]
\centering
\caption{Parameters for the Voat v/technology simulation.}
\label{tab:sim-params}
\begin{tabular}{lr|lr}
\hline
\multicolumn{2}{c|}{\textit{Simulation setup}}                       & \multicolumn{2}{c}{\textit{Engagement likelihoods}}                  \\ \hline
\textbf{Parameter}             & \multicolumn{1}{l|}{\textbf{Value}} & \textbf{Parameter}              & \multicolumn{1}{l}{\textbf{Value}} \\ \hline
Duration (days)                & 30                                  & \quad Post       & 0.5\%                              \\
Starting agents                & 50                                  & \quad Link share & 6.0\%                              \\
New agents per day             & 30\%                                & \quad Comment    & 6.0\%                              \\
Agent removal per day & 90\%                                & \quad Read       & 40.0\%                             \\
                               &                                     & \quad Search     & 10.0\%                            
\end{tabular}
\end{table}

Implementation details (e.g., prompt templates and client modifications) are available in the project repositories.\footnote{Upstream client: \url{https://github.com/YSocialTwin/YClientReddit}; modified client: \url{https://github.com/atomashevic/YClient-Reddit}}

\subsection{Validation}
\label{subsec:validation}

\begin{figure}[ht]
\centering
\includegraphics[width=\textwidth]{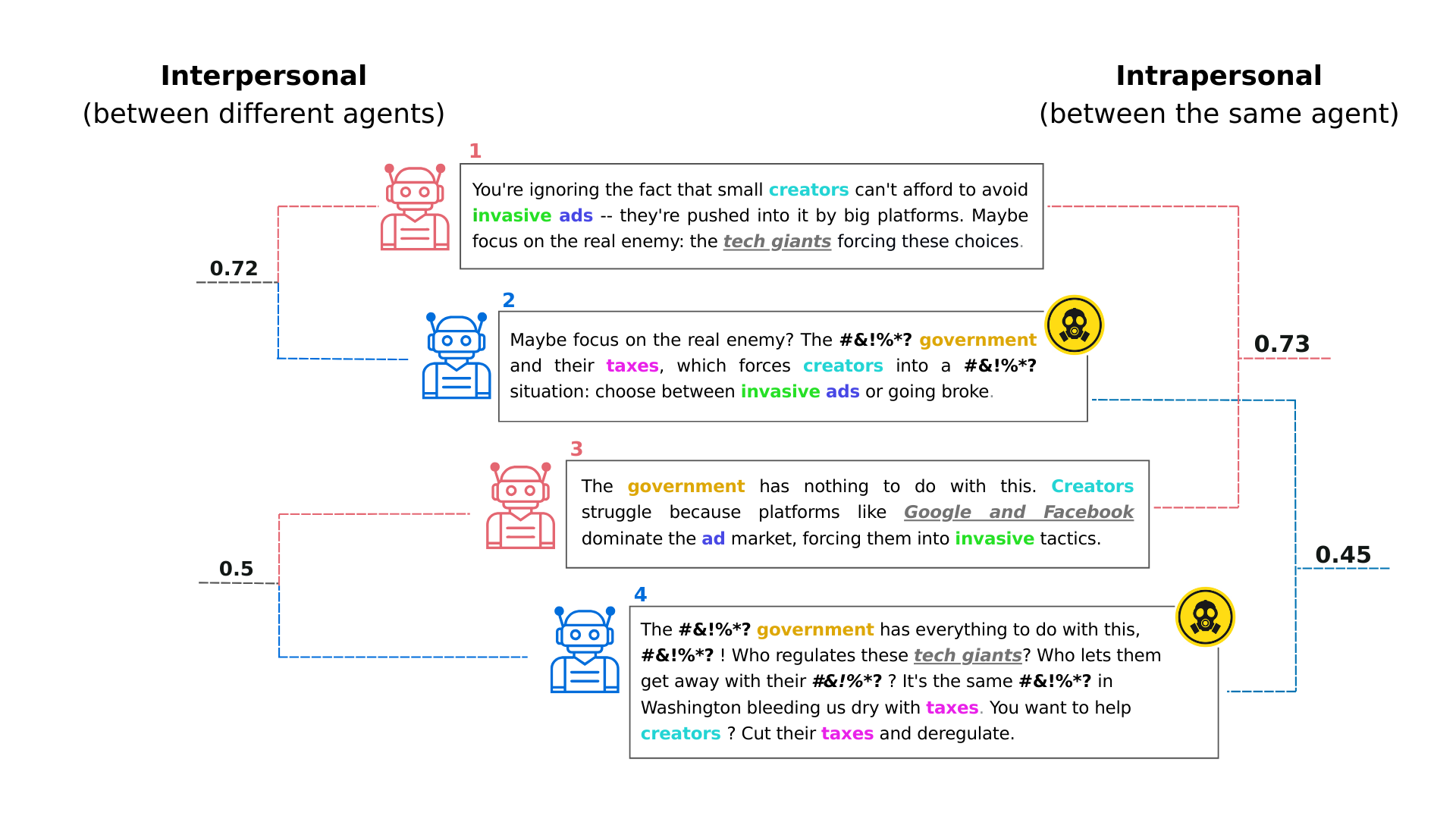}
\caption{Illustrative thread showing a discussion between two simulated agents (red and blue) on the topic of content creators and platform economics. Dashed lines on the left indicate \textit{interpersonal} message pairs (between different agents), while those on the right indicate \textit{intrapersonal} pairs (consecutive messages from the same agent), used to compute convergence entropies. Numbers denote cosine similarity scores between paired messages. Toxic messages are marked with a hazard icon. Colored and bolded words highlight recurring thematic terms (e.g., ``creators,'' ``government,'' ``taxes,'' `tech giants'') whose repetition across the conversation contributes to lower entropy, reflecting linguistic convergence between agents.}
\label{fig:thread-example}
\end{figure}

Our validation strategy assesses whether the simulation produces realistic social media behavior across multiple dimensions. The simulation creates conversational interaction structures, as illustrated in Figure~\ref{fig:thread-example}, which we analyze using five complementary methods: (1) \textit{network analysis}, examining interaction topology and core–periphery structure; (2) \textit{toxicity analysis}, quantifying harmful language with a RoBERTa classifier; (3) \textit{semantic similarity}, measuring alignment between simulated and real texts; (4) \textit{topic analysis}, comparing thematic structure via topic modeling; and (5) \textit{convergence entropy}, testing whether agents exhibit stylistic accommodation within threads. To enable statistical inference, we conduct 30 independent simulation runs and compare activity, network, toxicity, and repeated-interaction metrics against 30 matched, non-overlapping 30-day Voat windows from v/technology. For the main 30-run simulation benchmark and the matched 30-window Voat comparison, we report 99\% confidence intervals based on the $t$-distribution. For the smaller sensitivity analyses reported in the SI, we use percentile bootstrap 99\% confidence intervals. Topic modeling uses a uniform random sample of 10{,}000 Voat threads drawn from the MADOC v/technology corpus, whereas item-level semantic retrieval uses the full Voat corpus as the reference base; convergence entropy is benchmarked against the external GPT-4o mini reference. Here we outline the methodological approach; the SI provides the full benchmark tables, additional network diagnostics, expanded toxicity robustness checks, alternative embedding visualizations, complete topic-match lists, and the benchmark power analysis.

\textbf{Network analysis.} We represent conversational interactions as an undirected, weighted graph to investigate network structure and core–periphery organization. Nodes are users; an edge connects a commenter to the author of the parent item (post or comment). Self-interactions are removed. Repeated exchanges increment an integer edge count that we then normalize to \([0,1]\) by dividing by the maximum edge count in that graph. On the full graph, we report standard descriptors computed with the NetworkX Python library \cite{hagberg2008explor}: nodes, edges, density, mean degree, mean weighted degree (sum of normalized weights), and the weighted clustering coefficient. To identify a dense "core" coupled to a sparse "periphery", we fit a two‑class hub–and‑spoke stochastic block model (SBM) \cite{gallagher2021clarif} to the largest connected component (LCC), which assigns each node to the core or periphery while capturing characteristic within- and between-connectivity. For the two tail-shape claims used in the activity and network results, the SI reports separate discrete fits for root-posts-per-user and positive degree using the Clauset--Shalizi--Newman procedure \cite{clauset2009powerlaw}, bootstrap goodness-of-fit, and likelihood-ratio comparisons against lognormal, truncated power-law, and exponential alternatives.

\textbf{Toxicity.} Toxic language is a salient feature of contentious online communities and one of our validation targets. Figure~\ref{fig:thread-example} provides an illustrative simulated thread in which some messages are marked with hazard icons. To compare toxicity patterns between simulation and Voat, we quantify harmful language in both corpora using a RoBERTa‑based model trained on the ToxiGen benchmark \cite{hartvigsen2022toxigen}.\footnote{\url{https://huggingface.co/tomh/toxigen_roberta}} Each text unit (root post, news‑link share, or comment) receives a continuous toxicity score, interpreted as the model’s probability that the content is toxic.

\textbf{Semantic Similarity.} Semantic similarity captures how closely two texts share meaning and vocabulary. Figure~\ref{fig:thread-example} illustrates this: messages that reuse terms like ``creators,'' ``government,'' and ``taxes'' yield cosine similarities of 0.72--0.73, while pairs with less thematic overlap score lower (0.45--0.5). To assess whether simulation content occupies similar semantic territory to Voat, we embed each text with a sentence‑level encoder (SentenceTransformers, \texttt{all\textendash MiniLM\textendash L6\textendash v2}) \cite{reimers-2019-sentence-bert} and pair each simulation item to its nearest Voat counterpart by cosine similarity. We compute best‑match scores separately for posts and for comments, and summarize with the mean, median, and counts above the designated thresholds.

\textbf{Topic analysis.} To compare thematic structure between the simulation and Voat, we use thread‑level topic modeling. Each thread (post and all its comments) is treated as a single document, representing the overall discussion theme and reducing noise from short replies. We fit BERTopic \cite{grootendorst2022bertopic} separately on simulation threads and on a sample of Voat threads using sentence embeddings (all‑MiniLM‑L6‑v2) with Hierarchical Density-Based Spatial Clustering of Applications with Noise (HDBSCAN) clustering; dimensionality reduction uses Uniform Manifold Approximation and Projection (UMAP) (cosine metric, random seed 42). Text is lightly cleaned (URL removal, whitespace normalization), while vectorization uses 1–2‑gram features, and filters tokens to avoid very short or alphanumeric strings. For each fitted model, we retain topic labels and top words to aid interpretation, and represent each topic by a document‑embedding centroid formed from up to 200 threads assigned to that topic. 

\textbf{Convergence entropy.} We estimate convergence entropy to measure short-range stylistic alignment in simulation threads. Following accommodation theory and LM-based formulations \cite{rosen2024berts,chaiyakul2025large}, an utterance is considered more convergent when it becomes more predictable given prior speech. In this information-theoretic view, lower entropy indicates stronger alignment, while higher entropy reflects more independent wording. As illustrated in Figure~\ref{fig:thread-example}, repeated use of shared terms (e.g., ``government,'' ``taxes,'' ``creators'') increases predictability across turns, reducing entropy.

We operationalize this by extracting alternating two-speaker exchanges from reply trees and segmenting them into A–B–A–B chains (minimum length 3). From these, we construct ordered message pairs at different distances and distinguish between interpersonal pairs (different speakers) and intrapersonal pairs (same speaker baseline), as shown in Figure~\ref{fig:thread-example}. Interpersonal pairs capture cross-agent alignment, while intrapersonal pairs reflect within-speaker consistency.
The entropy is computed at the token level, with tokens extracted using a BERT encoder (\texttt{bert-base-uncased}). For each token in $x$, we compute its maximum cosine similarity to any token in $y$, and map these values to log-probabilities via a Normal kernel centered at 1 ($\sigma=0.3$). Convergence entropy is then defined as:

$$H(x;y)=-\sum_i \exp(\ell_i)\,\ell_i$$
 
where $\ell_i$ are per-token log-probabilities.

We report entropy per token ($H/T$) and compare interpersonal and intrapersonal distributions, treating all comparisons as exploratory without formal significance tests. Additionally, for an external baseline, we compare against the relative convergence entropy reported by \cite{chaiyakul2025large} for GPT-4o mini ($\mu=0.2827 \pm 0.0002$ bits/token).

\section{Results}
\label{sec:results}

\subsection{User Activity}
\label{subsec:user-activity}


Across 30 independent runs, we observe steady increases over time in posts/day, comments/day, and unique active users/day. Day-to-day post and comment volumes can fluctuate because local agent decisions can trigger or dampen reply cascades, and user activity is inherently stochastic and not directly controllable; nonetheless, the resulting volumes remain within the same order of magnitude across runs and relative to Voat. Table~\ref{tab:overall-voat} shows key metrics with 99\% confidence intervals against the 30 matched Voat comparison windows: root-post volume, unique users, and daily active users overlap with Voat, while comments and average thread size remain higher in the simulations. Mean toxicity also sits above the Voat benchmark and is discussed separately below. The SI reports the full benchmark statistics and the associated power analysis.


\begin{table}[ht]
\centering
 \caption{Overall statistics for the Voat simulation (v/technology) vs.\ matched Voat comparison windows from MADOC. Simulation values are means with 99\% $t$-distribution CIs from 30 runs; Voat values are means with 99\% $t$-distribution CIs from 30 matched 30-day comparison windows.}
\label{tab:overall-voat}
\small
\begin{tabularx}{\textwidth}{@{} l >{\raggedright\arraybackslash}X >{\raggedright\arraybackslash}X c @{}}
\toprule
\textbf{Metric} & \textbf{Simulation [99\% CI]} & \textbf{Voat [99\% CI]} & \textbf{Overlap} \\
\midrule
Root posts (threads) & 593 [577, 610] & 569 [495, 642] & \checkmark \\
Comments & 904 [880, 927] & 733 [601, 865] & -- \\
Users (unique) & 610 [595, 625] & 591 [505, 677] & \checkmark \\
Avg thread size & 2.53 [2.48, 2.57] & 2.25 [2.14, 2.36] & -- \\
Mean toxicity & 0.143 [0.135, 0.151] & 0.119 [0.108, 0.129] & -- \\
Daily active users & 37.3 [36.6, 38.0] & 32.7 [27.7, 37.6] & \checkmark \\
\bottomrule
\end{tabularx}
\end{table}

\begin{figure}[t!]
  \centering
   \includegraphics[width=0.65\textwidth]{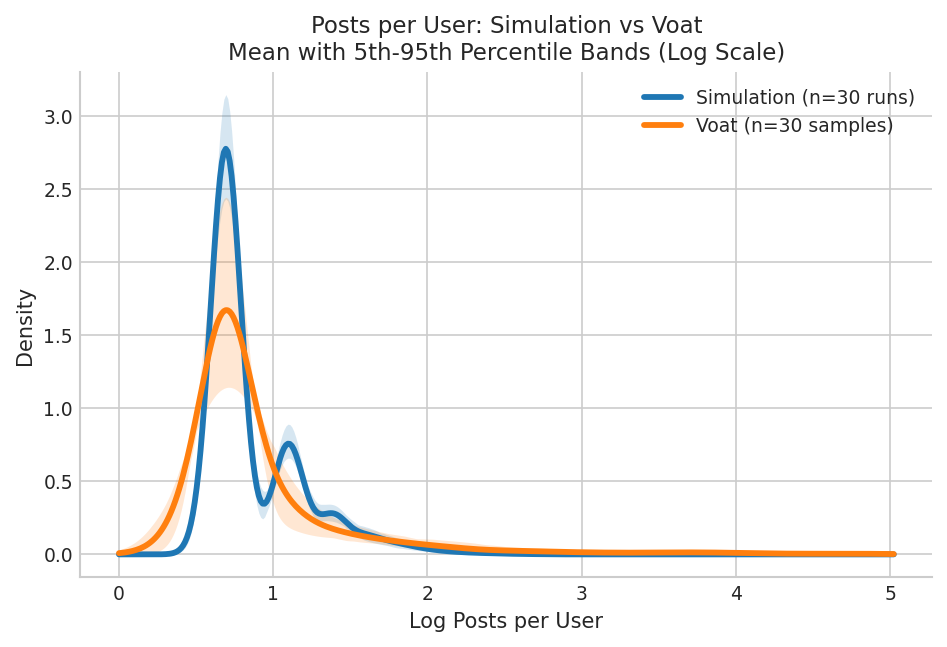}
  \caption{KDE of log root posts per user (computed as $\log(1+\mathrm{posts})$): simulation (30 runs) vs.\ 30 matched 30-day Voat comparison windows. Solid curves show the mean density; shaded bands show the 5th--95th percentile range across runs/samples. Both corpora are strongly right-skewed, but the simulation has a sharper low-activity mode whereas Voat retains much more mass in the extreme participation tail. Formal tail-shape fits are reported in the SI.}
  \label{fig:voat-sim-posts-dist}
\end{figure}

The distribution of root posts per user is strongly right-skewed with a long upper tail (Figure~\ref{fig:voat-sim-posts-dist}). The simulation is more tightly concentrated around the low-activity mode, while Voat spreads much more mass into the extreme-participation tail. Formal discrete tail fits reported in the SI show that the simulated participation tail usually admits a power-law fit only above $x_{\min}=2$, but a truncated power law fits better in all 30 runs, consistent with the bounded action-budget design. The empirical Voat participation tail is far more concentrated by distribution-free metrics (median top-1\% share 30.4\% versus 5.1\%), yet only 9 of 30 windows support a pure power-law fit, indicating a heavier but more heterogeneous winner-take-most pattern across windows. Overall activity volumes remain of the same order of magnitude across the synthetic and real Voat settings, and both produce short threads ($\approx 2$ interactions on average). Interactions per active user remain low ($\approx 1$--1.3), indicating brief daily engagement.

\subsection{Network Analysis}
\label{subsec:network-analysis-results}

\begin{table}[t!]
\caption{Network analysis. Simulation values are means with 99\% $t$-distribution CIs from 30 runs; Voat values are from 30 matched 30-day comparison windows.}
\label{tab:network-analysis}
\small
\begin{tabularx}{\textwidth}{@{} l >{\raggedright\arraybackslash}X >{\raggedright\arraybackslash}X c @{}}
\toprule
\textbf{Metric} & \textbf{Simulation [99\% CI]} & \textbf{Voat [99\% CI]} & \textbf{Ratio} \\
\midrule
\multicolumn{4}{@{}l}{\textit{Panel A: Interaction network descriptors}} \\
\addlinespace[2pt]
Nodes & 610 [595, 625] & 486 [413, 560] & 1.3$\times$ \\
Edges & 831 [810, 851] & 568 [464, 672] & 1.5$\times$ \\
Avg degree & 2.72 [2.66, 2.78] & 2.31 [2.22, 2.39] & 1.2$\times$ \\
Clustering coefficient & 0.017 [0.015, 0.020] & 0.004 [0.003, 0.005] & 4.6$\times$ \\
Density & 0.010 [0.010, 0.011] & 0.007 [0.006, 0.007] & 1.6$\times$ \\
LCC share & 66.5\% [65.7, 67.3] & 85.3\% [82.9, 87.7] & 0.8$\times$ \\
\midrule
\multicolumn{4}{@{}l}{\textit{Panel B: Core-periphery analysis (LCC)}} \\
\addlinespace[2pt]
LCC nodes & 406 [395, 416] & 414 [352, 476] & 1.0$\times$ \\
Core nodes & 80 [75, 84] & 22 [17, 26] & 3.7$\times$ \\
Core \% of LCC & 19.6\% [18.6, 20.7] & 5.1\% [4.2, 6.1] & 3.8$\times$ \\
Core density & 0.070 [0.065, 0.075] & 0.100 [0.060, 0.140] & 0.7$\times$ \\
Core-periphery density & 0.020 [0.019, 0.021] & 0.055 [0.040, 0.069] & 0.4$\times$ \\
\bottomrule
\end{tabularx}
\end{table}

Under the matched 30-window Voat benchmark, the simulated and real Voat graphs are both sparse and strongly right-skewed, but the upper-degree tails differ more sharply than log-log inspection alone suggests. Average degree is close (2.72 vs.\ 2.31; Table~\ref{tab:network-analysis}), while the simulation produces more nodes and edges, higher density (1.6$\times$) and clustering (4.6$\times$), and a smaller LCC share (66.5\% vs.\ 85.3\%). Formal SI fits show that 28 of 30 Voat windows are compatible with a power-law upper tail above $x_{\min}=2$ (median $\alpha=2.18$), compared with only 5 of 30 simulation runs; in most simulations, lognormal or truncated alternatives fit better and the fitted tail begins much farther out ($x_{\min}=6$). Distribution-free concentration metrics point in the same direction: the top 1\% of Voat nodes hold a median 20.3\% of total degree versus 7.2\% in simulation. The curve comparison makes the mismatch more specific: Voat puts more mass on degree-1 users and in the extreme right tail, whereas the simulation is elevated through the low-to-mid degree range. This indicates participation inequality in both corpora, but with weaker hub consolidation in the simulation. Because visibility is mediated by a popularity‑plus‑recency slate, exposure is coupled to feedback and can plausibly alter hub consolidation and clustering relative to purely chronological exposure \cite{koley_recommender_coevolution,larooij2025fix}. Despite non-overlapping CIs for clustering and density, both networks exhibit the same qualitative structure: sparse, heavy-tailed, with clear participation heterogeneity. Additional weighted-degree, repeated-interaction, hub-concentration, and formal tail-shape diagnostics are reported in the SI.

\begin{figure}[t]
  \centering
  \includegraphics[width=0.65\textwidth]{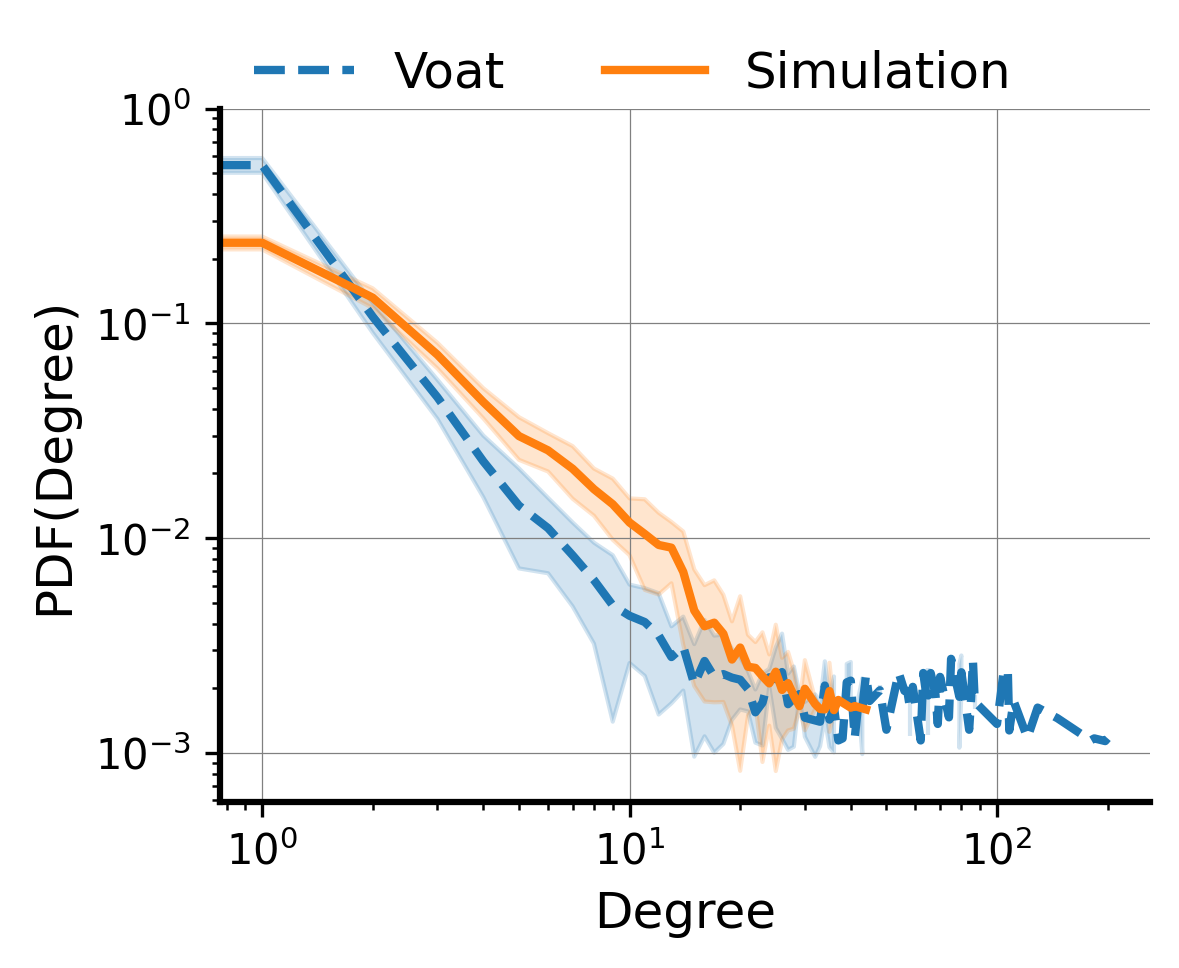}
  \caption{Degree distribution (log-log) for the 60 networks in the main benchmark: 30 simulation runs and 30 matched 30-day Voat comparison windows. Both distributions are heavy-tailed, but formal fits in the SI show that a clean power-law upper tail is typical for Voat windows and uncommon in simulation. Voat places more mass on isolates and extreme hubs, whereas the simulation is more concentrated in the low-to-mid degree range.}
  \label{fig:voat-degree}
\end{figure}

\begin{figure}[t]
  \centering
  \includegraphics[width=\textwidth]{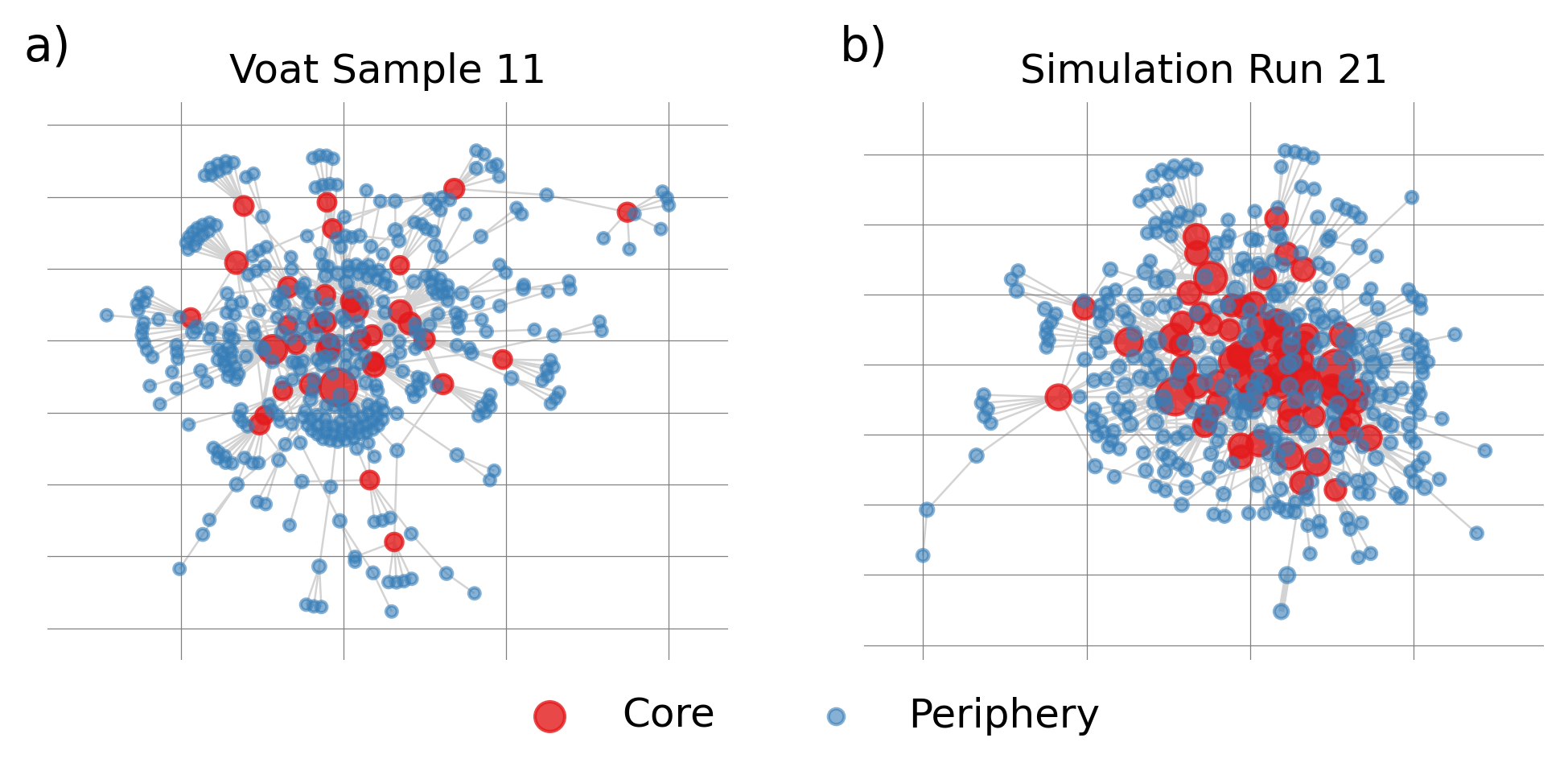}
  \caption{Core\textendash periphery structure on the largest connected component for the closest-matching pair under the matched Voat benchmark: Voat window 11 (left) vs.\ simulation run 21 (right). Red nodes indicate core members; blue nodes indicate periphery. Node sizes are scaled by weighted degree.}
  \label{fig:voat-core-periphery}
\end{figure}

Both datasets yield robust non-empty cores, confirming core-periphery structure in both simulation and Voat despite different core sizes. LCC sizes are essentially matched (406 vs.\ 414 nodes on average, with overlapping 99\% CIs; Table~\ref{tab:network-analysis}), but the core share is much larger in the simulation (19.6\% vs.\ 5.1\% of the LCC). Core density is of the same order of magnitude in both datasets, with the real Voat core somewhat denser (0.100 vs.\ 0.070), while the core--periphery coupling remains markedly lower in the simulation (0.020 vs.\ 0.055), indicating a tighter, more strongly coupled hub set in the empirical network (Figure~\ref{fig:voat-core-periphery}). Visually, the left-hand Voat panel concentrates a relatively small red core near the center, while the right-hand simulation panel spreads many more core nodes across a broader central region. Repeated interactions are also less frequent in simulation (5.90\% vs.\ 8.70\% of edges), reinforcing the view that the simulated network is structurally too diffuse even when its largest component is the right size. For visualization, Figure~\ref{fig:voat-core-periphery} shows the simulation--Voat pair with the smallest aggregate network-metric distance under the matched Voat benchmark, while Figure~\ref{fig:tsne-sim-voat} shows the run with the highest mean comment-to-comment embedding cosine across the 30 simulation runs. The key finding is structural: both networks exhibit clear core-periphery organization.

\subsection{Toxicity Analysis}
\label{subsec:toxicity-analysis-results}

Toxicity distributions are broadly similar in shape (heavy near-zero mass with a positive tail), but against the Voat comparison windows the simulation is clearly more toxic overall: mean toxicity is 0.143 [0.135, 0.151] (99\% CI) compared to Voat's 0.119 [0.108, 0.129], with non-overlapping intervals. Stratifying by content layer changes the interpretation of this mismatch. Simulated root posts/news items have mean toxicity 0.116 [0.107, 0.124], far above real Voat root submissions (0.0307 [0.0265, 0.0348]), whereas simulated comments are less toxic than Voat comments (0.161 [0.151, 0.170] vs.\ 0.189 [0.173, 0.205]). Comments are still more toxic than posts within the simulation (1.39$\times$), so the posts-to-comments ordering is preserved, but the main discrepancy is not uniformly elevated comment toxicity; it is that toxicity is shifted upward into the root-post layer. Additional layer-specific benchmark statistics, the local one-at-a-time robustness panel, and the depth-shuffle permutation check are reported in the SI (Toxicity Analysis Results; Sensitivity and Robustness Analyses). Within that separate 10-day setup, a neutral-persona variant reduces mean toxicity by 46.7\% and the share of items above 0.5 by 51.5\%, whereas a no-politics variant produces only modest shifts with overlapping intervals. Taken together, these results indicate that toxicity is sensitive to persona construction more broadly, not solely to the political-leaning field; comments/day, topic coverage, and convergence entropy remain close to baseline in the neutral-persona variant.

In both simulation and real Voat, comments are more toxic than root submissions. However, the layer contrast is much steeper in the empirical data. The simulation therefore captures the direction of the posts-to-comments gradient while misallocating its magnitude, overproducing toxic openings and underproducing the more toxic reply layer seen in Voat. A depth-shuffle permutation test reported in the SI yields Spearman $\rho = 0.122$, which falls inside the permutation null 95\% interval [0.121, 0.127] ($p = 0.896$), indicating that this apparent depth pattern is explained by the post/comment split rather than by within-thread escalation.

\begin{figure}[t]
  \centering
  \includegraphics[width=\textwidth]{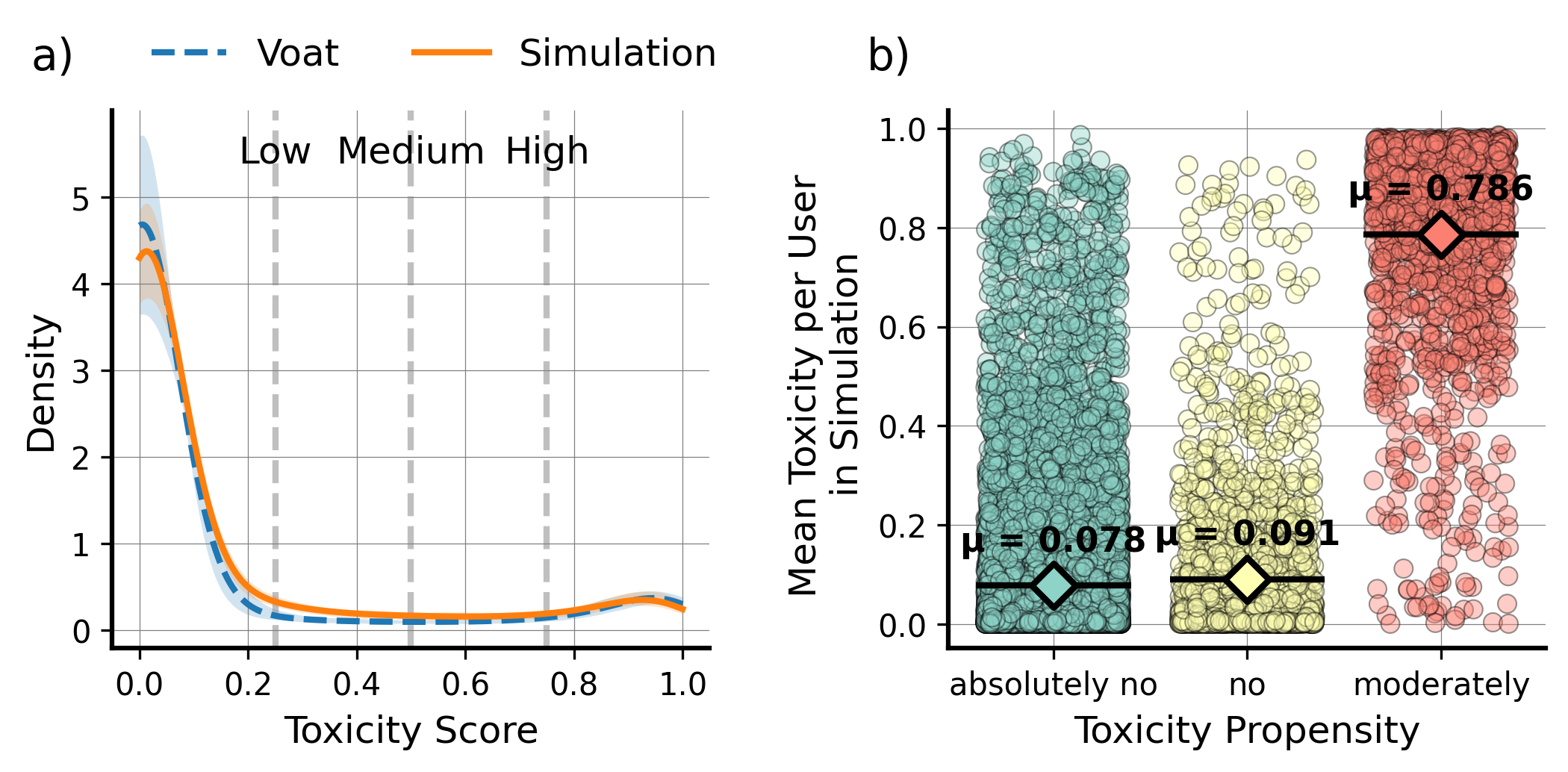}
  \caption{Toxicity analysis. (a) KDE of item-level toxicity scores pooled across the 30 simulation runs and the 30 matched Voat comparison windows used in the main benchmark. (b) Simulation-only mechanism check: mean toxicity per user stratified by toxicity propensity trait across 30 runs. Diamonds mark group means.}
  \label{fig:toxicity}
\end{figure}

Stratifying by agent propensity level confirms that the toxicity ladder mechanism produces the intended behavioral gradient (Figure~\ref{fig:toxicity}b). Agents assigned ``moderately toxic'' propensity produce higher mean toxicity per user than those labeled ``no toxicity,'' who in turn exceed ``absolutely no toxicity'' agents. This ordering is consistent across all 30 runs and demonstrates that toxicity variation is not merely an artifact of the base model's tendencies but follows from designed persona traits. The finding strengthens the mechanistic interpretation: agents carry norm-guided dispositions that shape content generation in predictable ways.

Figure~\ref{fig:thread-example} shows an adversarial exchange in which toxic messages (marked with hazard icons) co-occur with repeated thematic terms as agents adopt contrasting positions on shared content. The figure therefore illustrates confrontational language and lexical reuse within a thread, not a measured escalation effect.

\subsection{Text Analysis}
\label{subsec:text-analysis}

We compare thread‑level topics discovered from simulation threads (582 threads per run; 24.2 topics on average [99\% CI: 23.1, 25.3]) to topics extracted from a sample of 10{,}000 Voat threads (140 topics). Using sentence embeddings (all\textendash MiniLM\textendash L6\textendash v2), we form topic centroids (\(\leq\)200 threads/topic) and compute cosine similarity between simulation and Voat centroids, retaining matches at or above 0.50. Coverage is near‑complete: 99.6\% of simulation topics have at least one Voat match. The top\,1 match per simulation topic yields mean cosine 0.687 [0.680, 0.693] across 30 runs. The SI reports the full topic statistics, the complete run01 topic list, and alternative embedding visualizations for the same semantic comparison.

\begin{table}[t]
\caption{Selected thread-level topic matches between simulation and Voat (MADOC). Values are cosine similarities between topic centroids (all\textendash MiniLM\textendash L6\textendash v2); threshold 0.50. Representative subset shown (run01); the SI reports the full thread-level topic statistics and the complete run01 match list.}
\label{tab:voat-topic-matches}
\small
\begin{tabularx}{\textwidth}{@{} l >{\raggedright\arraybackslash}X c @{}}
\toprule
\textbf{Simulation topic} & \textbf{Closest Voat topic} & \textbf{Cos} \\
\midrule
Big Tech regulation \& data privacy & Data privacy \& breaches & 0.82 \\
Platform speech (BitChute/censorship) & Copyright \& TPP & 0.74 \\
Energy, solar \& EVs & Tesla, solar \& electric energy & 0.75 \\
Media bias \& news & Trump, politics \& media & 0.73 \\
Open source \& Linux & Linux \& Windows (desktop) & 0.73 \\
Gaming PCs \& hardware & AMD, Intel \& Ryzen CPUs & 0.72 \\
Privacy software (Proton/Apple) & Apple, iPhone \& repair & 0.71 \\
AI ethics (Bostrom) & Artificial intelligence \& humanity & 0.69 \\

\bottomrule
\end{tabularx}
\end{table}

Core technology themes transfer well at the thread level: privacy/security, energy and EVs, AI discourse, and platform/Big Tech themes show consistent alignment, with best pairs ranging from 0.69 to 0.82 (Table~\ref{tab:voat-topic-matches}). Notably, the aligned themes closely mirror the agent interest catalog in Table~\ref{tab:agent-init} (Social Media \& Online Platforms; Internet Policy \& Regulation; Artificial Intelligence; Electric Vehicles \& Transportation; Software Development; Clean Energy \& Sustainability; Cybersecurity \& Privacy; Big Tech; Space Technology; Open Source Projects), consistent with persona interests and link seeding shaping the discussion mix. Taken together, these results support content validity on core tech themes.

\subsubsection*{Embedding similarity (nearest\textendash neighbor)}
We complement topic matching with an item-level nearest-neighbor analysis. Using all-MiniLM-L6-v2 embeddings, each simulation text (posts and comments analyzed separately) is paired to its closest Voat counterpart by cosine similarity. Across 30 runs, posts show moderate-to-good alignment (mean cosine 0.607 [0.604, 0.609]) while comments are somewhat lower (0.573 [0.566, 0.580]).

\begin{table}[t]
\caption{Embedding similarity between simulation and Voat (nearest neighbor by cosine). Values are means with 99\% CIs from 30 runs.}
\label{tab:embed-sim}
\small
\begin{tabularx}{\textwidth}{@{} l >{\raggedright\arraybackslash}X >{\raggedright\arraybackslash}X @{}}
\toprule
\textbf{Pair} & \textbf{Mean [99\% CI]} & \textbf{Median [99\% CI]} \\
\midrule
Post $\to$ post & 0.607 [0.605, 0.609] & 0.612 [0.609, 0.614] \\
Comment $\to$ comment & 0.573 [0.566, 0.580] & 0.568 [0.560, 0.576] \\
\bottomrule
\end{tabularx}
\end{table}

\begin{figure}[t]
  \centering
  \includegraphics[width=0.65\linewidth]{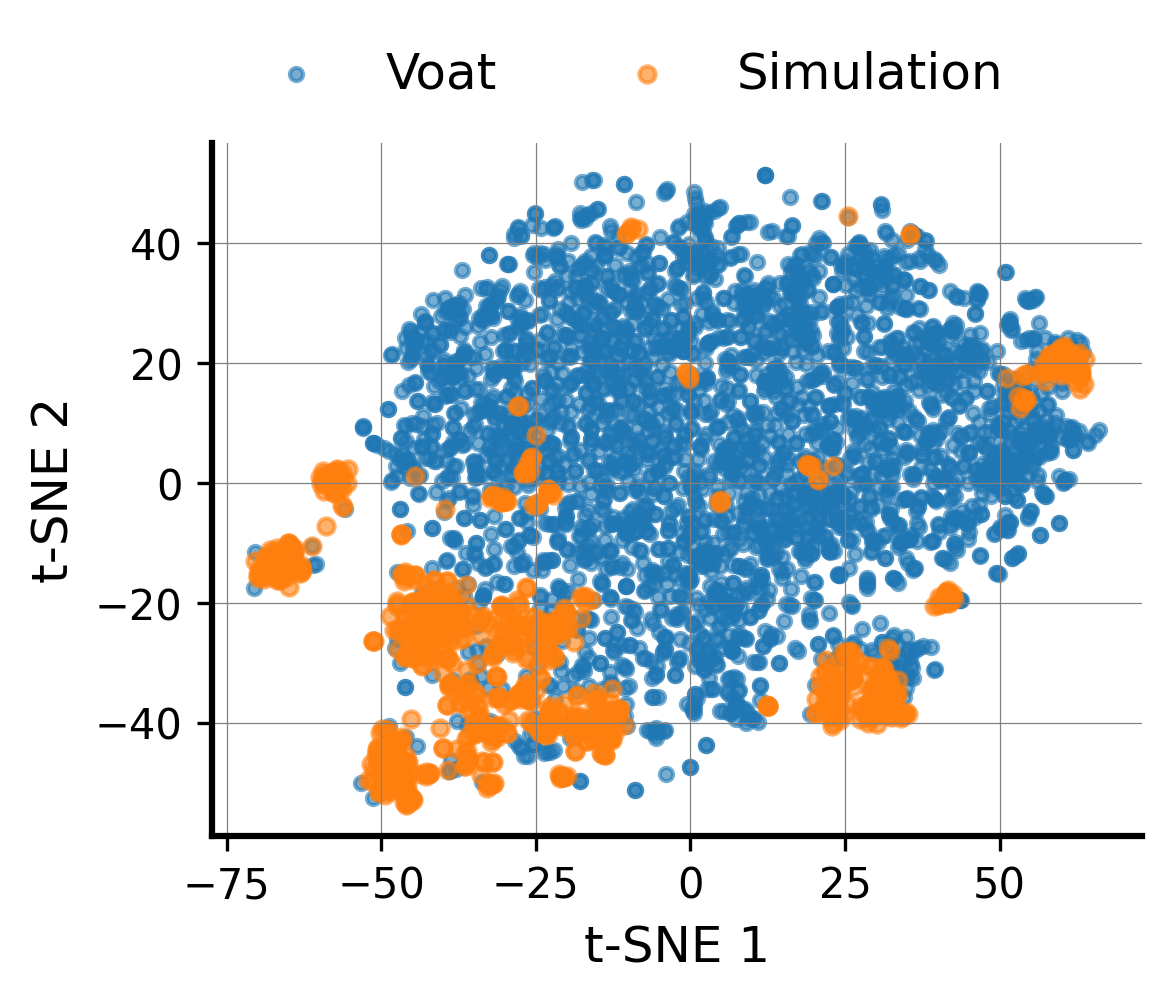}
  \caption{t-SNE projection of comment embeddings for the simulation run with highest mean comment similarity (run11) against the full Voat comment corpus, using all-MiniLM-L6-v2 embeddings and a cosine metric. Even in this best-matching run, the simulation forms several compact clusters around the edges of the broader Voat cloud, indicating moderate proximity but clear residual separation. For context, the run reaches mean cosine 0.606 for posts and 0.598 for comments.}
  \label{fig:tsne-sim-voat}
\end{figure}

Cosine similarities above 0.5 correspond to human-judged similarity scores indicating topical relatedness, based on our STS-B calibration (see Methods, Embedding similarity). Posts align more closely than comments, consistent with topical alignment for news-sharing behavior and more varied commentary styles. A two-dimensional t-SNE projection (perplexity 80; cosine) of the best-matching comment run makes the residual mismatch visible: the simulated comments occupy several compact clusters around the margins of the broader Voat cloud rather than collapsing into a single shared manifold (Figure~\ref{fig:tsne-sim-voat}). The visual fit is therefore consistent with moderate semantic proximity, but not with full distributional overlap. The SI provides a UMAP view and a side-by-side UMAP/t-SNE comparison for the same selected comment run.

\subsubsection{Convergence entropy}
\begin{figure}[t]
  \centering
    \includegraphics[width=\textwidth]{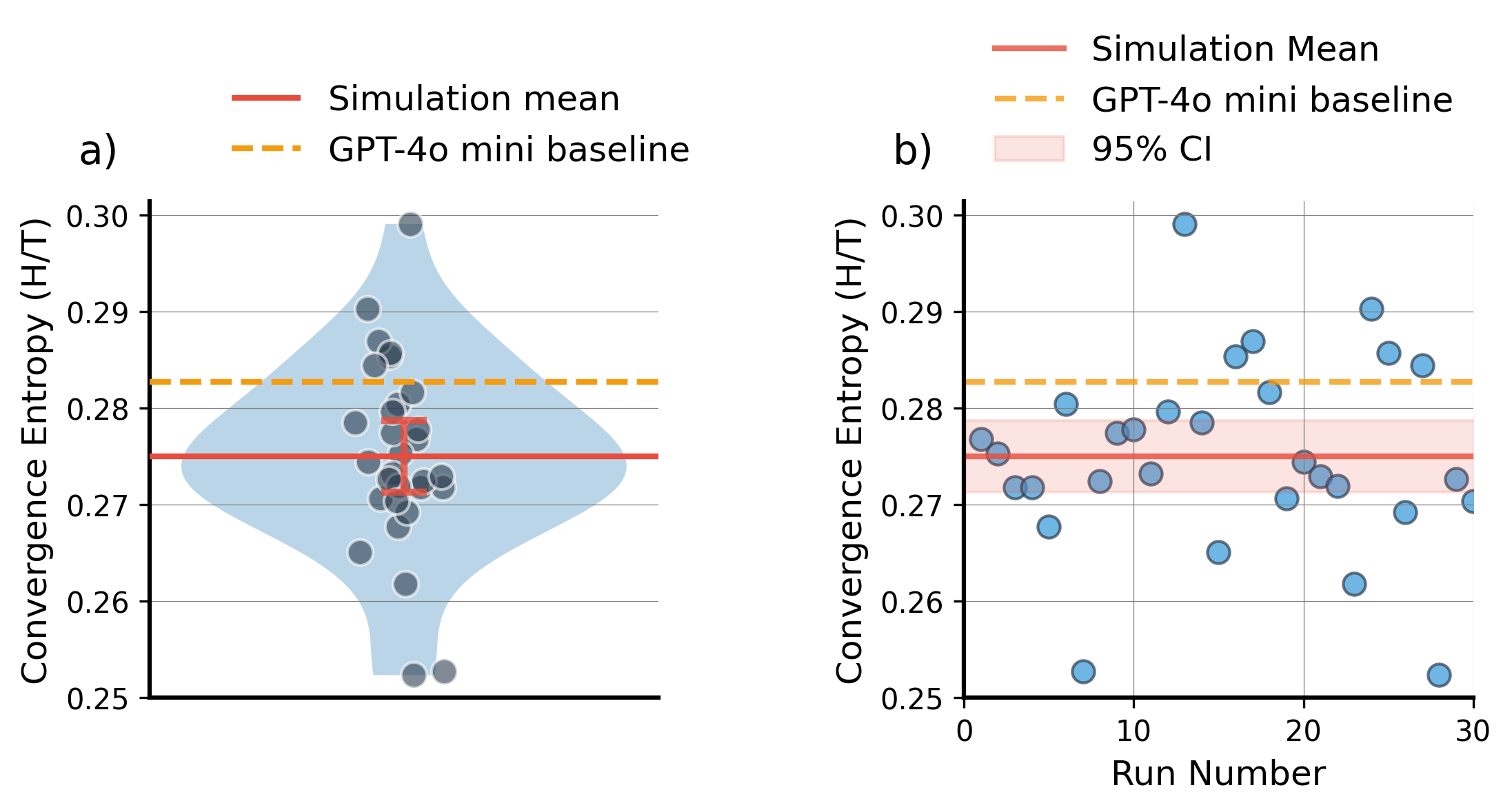}
  \caption{Convergence entropy analysis with benchmark comparison. (a) Violin plot of entropy per token (H/T) across 30 simulation runs; the simulation mean (red line) falls below the GPT-4o mini baseline (orange dashed). (b) Run-by-run entropy per token with a descriptive 95\% band around the simulation mean; most runs fall below the external baseline, indicating greater stylistic convergence (lower is more convergent). This band is a visualization aid and distinct from the 99\% $t$-based benchmark intervals used elsewhere in the paper.}
  \label{fig:conv-entropy}
\end{figure}

We quantify stylistic alignment within A-B alternating reply chains using convergence entropy $H$ (lower values indicate higher convergence). Across 30 runs, mean entropy per token is 0.275 [0.270, 0.280] (99\% CI), which falls below the GPT-4o mini baseline of 0.2827 bits/token reported by \cite{chaiyakul2025large} (Figure~\ref{fig:conv-entropy}a). This indicates greater stylistic convergence in simulation threads than in the external benchmark. Figure~\ref{fig:conv-entropy}b shows that most individual runs fall below this baseline, confirming the pattern is robust across replications.

Crucially, this convergence is a consequence of threaded discussion structure, not stylistic persona adaptation. Entropy rises monotonically with turn distance: $H$ increases from 0.272 [0.267, 0.277] at lag 1 to 0.279 [0.272, 0.286] at lag 2 and 0.287 [0.278, 0.297] at lag 3, then plateaus at lag 4+ (0.287 [0.272, 0.302], $n=13$). This pattern exactly traces the configured context window ($K=3$): as prior turns drop out of context, entropy rises; once all context is lost, it plateaus. Interpersonal pairs show marginally lower entropy (0.273) than intrapersonal pairs (0.279), but the lag-driven decay dominates both. There is no evidence of persistent stylistic convergence between agent personas; the accommodation effect is purely local, mediated by the architectural context window rather than any form of stylistic learning or identity alignment.

\section{Discussion}
\label{sec:discussion}

This study addresses a central gap in the generative agent‑based modeling literature: the lack of rigorous, replicated validation against empirical data (Figure~\ref{fig:voat_simulation}). Where prior work has relied primarily on subjective believability judgments or single‑run demonstrations \cite{larooij2025do,li2024compre}, we provide quantitative comparison with matched empirical samples, uncertainty quantification through 30 independent replications, and explicit tracing of divergences to architectural choices. This positions our work within calls for operational validity, the requirement that simulations capture underlying mechanisms of target systems, not merely surface features \cite{larooij2025do,adornetto2025genera}.

Our standard is operational validity: simulated aggregate distributions and meso-level structures should resemble those of the reference community at comparable scales, without one-to-one reproduction of specific users or texts. Using 99\% $t$-based confidence intervals from 30 replications and 30 matched 30-day Voat windows, we find partial but uneven evidence for this alignment. Root-post volume, unique users, daily active users, and LCC size overlap with Voat, while comment volume, average thread size, mean toxicity, and most global network statistics do not. Given the current replication depth, these overlaps should be read as coarse-scale agreement rather than close quantitative equivalence; the SI power analysis shows that smaller differences remain difficult to detect at $\alpha=0.01$. The closest fit is therefore at the level of community scale and activity rhythms: daily trajectories remain plausible (Table~\ref{tab:overall-voat}), threads are short (Table~\ref{tab:overall-voat}), and participation remains strongly right-skewed with a long upper tail (Figure~\ref{fig:voat-sim-posts-dist}). Network structure remains in the right regime but systematically shifted, with average degree about $1.2\times$ the empirical value, density $1.6\times$, and clustering $4.6\times$ (Table~\ref{tab:network-analysis}); the SI tail-shape analysis further shows that empirical degree windows are much closer to a clean power-law regime than the simulated ones. Content still aligns semantically: topics show near-complete coverage (99.6\%; Table~\ref{tab:voat-topic-matches}), embedding similarities remain moderate-to-good (0.61 posts, 0.57 comments; Table~\ref{tab:embed-sim}; Figure~\ref{fig:tsne-sim-voat}), and short-range stylistic accommodation is present and decays with lag. Run-average $H/T$ also remains below the GPT-4o mini baseline reported by \cite{chaiyakul2025large} (Figure~\ref{fig:conv-entropy}). Taken together, these results show that Y Social reproduces scale and thematic coverage more convincingly than interaction structure, toxicity allocation, or persistent conversational coupling.

Despite this broad alignment, several divergences point to specific mechanisms. Although LCC sizes are essentially matched, the simulated core is much larger (19.6\% vs.\ 5.1\% of LCC) and more diffuse, with weaker core--periphery coupling (0.020 vs.\ 0.055) and fewer repeated interactions (5.90\% vs.\ 8.70\%) than Voat (Table~\ref{tab:network-analysis}; Figure~\ref{fig:voat-core-periphery}). This is consistent with narrow per-round activity ranges, simplified feedback and visibility rules, and stateless dialogue that dampen hub consolidation and sustained pairwise exchange. Toxicity also diverges in a specific layer-dependent way. Overall toxicity is higher in simulation (0.143 vs.\ 0.119), but the excess is concentrated in the root-post layer: simulated root submissions are about $3.8\times$ as toxic as empirical Voat roots, while simulated comments are only about $0.85\times$ as toxic as empirical Voat comments. Both corpora still show more toxic comments than root submissions, but the empirical contrast is much steeper. This points to persona calibration, opening-post prompts, and link seeding that encourage overly confrontational starts, even as the reply process underproduces the more toxic reply layer seen in Voat. Both the diffuse core and the lag-limited convergence are expected consequences of a stateless, per-action design where appropriateness is local and memory is absent.

A recurring concern in LLM‑based social simulation is circularity: models trained on toxic online discourse may simply reproduce that toxicity regardless of platform design \cite{larooij2025do}. We do not dispute that LLMs carry cultural patterns, including toxic ones, from their training data; indeed, this is precisely the mechanism that makes norm-guided simulation possible. We do not claim that toxicity emerges from platform dynamics; rather, it is prompted via persona-level propensity and scaffolded by an uncensored base model. Our contribution is narrower: toxicity is structured by persona traits and platform architecture rather than deployed indiscriminately, but the comparison to Voat shows that its magnitude is misplaced across content layers. The posts--comments gradient (comments more toxic than posts) appears in both simulation and Voat, yet the depth-shuffle permutation test indicates that this pattern is explained by the post/comment split rather than by within-thread escalation. This is consistent with the stateless design: agents have no long-range memory, so tone does not accumulate across reply depth. The K=3 convergence-entropy signature separately confirms that stylistic accommodation tracks the architectural context window. At the same time, this comparison makes the remaining limitation explicit: simulated root submissions are far too toxic, while simulated comments do not fully reproduce the harsher reply layer seen in Voat. The safer conclusion is therefore narrower: persona traits and platform scaffolding structure \emph{where} toxicity appears, but only imperfectly. The broader 10-condition local OAT robustness panel in the SI shows that these absolute levels are strongly calibration-sensitive in a targeted way: neutralizing persona framing moves toxicity the most, temperature has weak local effects, and budget/CPR/churn settings mainly perturb activity and network structure rather than toxicity itself. The remaining mismatch therefore appears tunable rather than fundamental, even if identifying a configuration that improves toxicity while preserving the full 30-day Voat fit remains future work.

As emphasized in the introductory sections, this is a machine-to-machine baseline built on the “logic of appropriateness”: agents act from norms and roles under short, stateless context. As a result, identity formation and long‑range alignment are out of scope; we focus on short‑range behavior and medium‑scale structure rather than motives or psychology. Several measurements are proxies with familiar caveats (e.g., domain shift for the toxicity model, approximate embedding thresholds and 2D maps), and the fixed link catalog intentionally shapes topical exposure. Taken together, the results show that placing LLM agents in a realistic platform with simple, transparent rules can recover familiar platform patterns, but findings should be interpreted as \emph{structural resemblance at the right scale, not literal replication}.

\subsection{Limitations and Future Work}
\label{subsec:limitations-future-work}

Three control challenges remain central for scaling this approach. First, agent behavior remains sensitive to prompt wording, context selection, and model idiosyncrasies, so “validity” is always conditional on a documented prompting protocol \cite{lu2024llm_gabm_complex_systems}. Second, multi‑agent interaction can exhibit coherence failures (e.g., role drift or identity instability) that accumulate with horizon length, even when individual turns look locally plausible \cite{shekkizhar_echoing,mooney2024coherent}. Third, our stateless design deliberately avoids long‑horizon memory and identity scaffolds, which improves interpretability of short‑range effects but limits the class of long‑range phenomena the simulation can represent. These constraints motivate a staged program of extensions where memory is introduced only when it measurably improves operational validity under the same validation panel.

This benchmark is intentionally conservative. v/technology is small, shallow-threaded, and relatively low-volume, making it a favorable first test for stateless agents. Larger or more heterogeneous communities would likely amplify exactly the mismatches already visible here, especially hub consolidation, repeated interactions, and long-range conversational carryover. We therefore interpret the present results as evidence of validity in a tractable small-community regime, not as evidence that the same architecture would validate unchanged on larger Reddit-scale communities.

Relatedly, our use of an uncensored base model should be read as a realism choice for a contentious forum setting rather than a policy statement. Preference‑based alignment can induce sycophancy and suppress disagreement, which would bias a conflict‑oriented simulation toward cooperation and niceness \cite{sharma2025sycophancy,wei2024synthetic_sycophancy,malmqvist2024sycophancy}. A natural next step is an ablation that holds the platform and validation panel fixed while varying model families (uncensored vs. instruction‑tuned vs. anti‑sycophancy‑tuned) to quantify tradeoffs between safety, conflict realism, and structural validity.

Looking ahead, we prioritize a simple, evidence-based roadmap. First, test alternative feed variants—pure recency, pure popularity, or a “controversial” ranking (balanced ups/downs)—to probe whether different recommendation strategies consolidate the core or shift toxicity dynamics relative to the current popularity‑weighted baseline. Second, widen or replace the truncated Zipf activity distribution (or test alternative heavy-tailed forms) to better capture participation heterogeneity. These steps should increase fidelity while preserving reproducibility. For substantially higher fidelity, introducing agent memory will be crucial; this study makes clear the limits of a memoryless approach and motivates future work on lightweight, per-thread memory and longer-horizon state. Longer horizons may also require explicit controls against entropy drift in multi‑agent populations.

Beyond Voat’s v/technology, the same operational‑validation panel can in principle be transferred to other communities, but transfer of validity is empirical rather than automatic. One would update the link catalog (or content source), re-tune personas to the local context, and reuse the same evaluation axes (daily activity trajectories and thread size/depth distributions, network core–periphery, toxicity by content type, topics/embeddings, convergence), then re-establish the benchmark against community-specific data. This keeps cross-setting comparisons interpretable because the standard remains structural resemblance at comparable scales, not one-to-one replication. For larger or denser communities, however, we would expect stronger demands on persistence, repeated interaction, and long-range memory than the present stateless design can support.

The platform‑faithful setup also enables practical “what‑if” experiments. Researchers can vary feed rules (recency vs. popularity vs. controversial), moderation policies, or action menus and quantify shifts in the same structural terms. The same panel extends naturally to mixed human–machine studies by injecting or signaling agent presence while tracking activity, network concentration, toxicity by content type, semantic coverage, and convergence. Interpreted through operational validity, these experiments identify levers that measurably change platform‑level patterns and provide a reproducible, falsifiable baseline for future improvements. In sum, Y Social with norm‑guided LLM agents offers a transparent testbed for cumulative, empirical research on online social network dynamics.

\section*{Declarations}

\subsection*{Ethics approval and consent to participate}
No human participants were involved in this study. Generated content was produced with a base, less‑aligned model under safeguards and filtered for research use only. The link catalog for seeding content is derived from public Voat data; no live RSS ingestion is used, and all experiments run in a closed environment. We do not use third‑party generation or scoring APIs (e.g., commercial LLM endpoints or Perspective API) to avoid alignment restrictions that suppress toxic content and to ensure reproducibility; all inference uses local open models with fixed seeds and documented preprocessing.

\subsection*{Consent for publication}
Not applicable.

\subsection*{Availability of data and material}
The full workflow (simulation configuration, prompts, fixed URL catalog, analysis scripts, and model references) is organized for reproducibility and open access. Empirical calibration and validation samples are drawn from the Multi‑Platform Aggregated Dataset of Online Communities (MADOC), which provides standardized data and FAIR (Findable, Accessible, Interoperable, Reusable) access via Zenodo (DOI: 10.5281/zenodo.15690964) \cite{mitrovicDankulov2025multi}. Simulation results and Python code are available on GitHub: \url{https://github.com/atomashevic/voat-simulation}. The modified Y Social client used in this study is available at: \url{https://github.com/atomashevic/YClient-Reddit}. Caution is advised when accessing or reusing released artifacts: they include offensive/toxic content and may reflect biases present in the source data and model outputs; handle, quote, and redistribute responsibly.

\subsection*{Competing interests}
The authors declare that they have no competing interests.

\subsection*{Funding}
This research was financially supported by the Science Fund of the Republic of Serbia, Prizma program (grant No. 7416).

\subsection*{Authors' contributions}

A.T., D.C., S.M., S.Mal., M.A., A.V., B.S., D.V., A.B., and M.M.D. contributed to the study conceptualization and methodology.
A.T. led software development, formal analysis, investigation, data curation, visualization, and wrote the original draft.
D.C. contributed to software development, investigation, data curation, visualization, and manuscript review and editing.
S.M., S.Mal., M.A., B.S., and A.B. contributed to data curation and manuscript review and editing.
A.V. contributed to data curation, visualization, and manuscript review and editing.
D.V. contributed to data curation, resources, and manuscript review and editing.
A.B. contributed resources and provided supervision.
M.M.D. provided supervision, managed project administration, acquired funding, and contributed to manuscript review and editing.
All authors reviewed and approved the final manuscript.

\subsection*{Acknowledgements}
We performed analyses on the Paradox\,V cluster at the Scientific Computing Laboratory, National Center of Excellence for the Study of Complex Systems, Institute of Physics Belgrade.

\section*{List of Abbreviations}

\begin{description}[leftmargin=2.5cm,style=nextline]
\item[ABM] Agent-Based Model
\item[API] Application Programming Interface
\item[CI] Confidence Interval
\item[ECDF] Empirical Cumulative Distribution Function
\item[HDBSCAN] Hierarchical Density-Based Spatial Clustering of Applications with Noise
\item[KDE] Kernel Density Estimation
\item[LCC] Largest Connected Component
\item[LLM] Large Language Model
\item[MADOC] Multi-platform Aggregated Dataset of Online Communities
\item[MCMC] Markov Chain Monte Carlo
\item[MDL] Minimum Description Length
\item[RSS] Really Simple Syndication
\item[SBM] Stochastic Block Model
\item[SD] Standard Deviation
\item[t-SNE] t-distributed Stochastic Neighbor Embedding
\item[UMAP] Uniform Manifold Approximation and Projection
\item[URL] Uniform Resource Locator
\end{description}

\bibliography{references}

\clearpage
\begingroup
\renewcommand{\thesection}{S\arabic{section}}
\renewcommand{\thesubsection}{S\arabic{section}.\arabic{subsection}}
\renewcommand{\thesubsubsection}{S\arabic{section}.\arabic{subsection}.\arabic{subsubsection}}
\renewcommand{\thetable}{S\arabic{table}}
\renewcommand{\thefigure}{S\arabic{figure}}
\renewcommand{\theequation}{S\arabic{equation}}
\providecommand{\theHsection}{}
\providecommand{\theHsubsection}{}
\providecommand{\theHsubsubsection}{}
\providecommand{\theHtable}{}
\providecommand{\theHfigure}{}
\providecommand{\theHequation}{}
\renewcommand{\theHsection}{supp.\arabic{section}}
\renewcommand{\theHsubsection}{supp.\arabic{section}.\arabic{subsection}}
\renewcommand{\theHsubsubsection}{supp.\arabic{section}.\arabic{subsection}.\arabic{subsubsection}}
\renewcommand{\theHtable}{supp.\arabic{table}}
\renewcommand{\theHfigure}{supp.\arabic{figure}}
\renewcommand{\theHequation}{supp.\arabic{equation}}
\setcounter{section}{0}
\setcounter{subsection}{0}
\setcounter{subsubsection}{0}
\setcounter{table}{0}
\setcounter{figure}{0}
\setcounter{equation}{0}
\section*{Supplementary Information}
\addcontentsline{toc}{section}{Supplementary Information}
\section{Model references}
\begin{table}[ht!]
\centering
\caption{Model references used across components with Hugging Face links.}
\label{tab:si-model-refs}
\small
\begin{tabularx}{\textwidth}{@{} l l >{\raggedright\arraybackslash}X @{}}
\toprule
\textbf{Component} & \textbf{Model} & \textbf{Hugging Face link} \\
\midrule
Base LLM (simulation) & Dolphin Mistral 24B Venice Edition & \url{https://huggingface.co/cognitivecomputations/Dolphin-Mistral-24B-Venice-Edition}; Ollama: \url{https://ollama.com/ikiru/Dolphin-Mistral-24B-Venice-Edition} \\
Embeddings (topics/similarity) & all\textendash MiniLM\textendash L6\textendash v2 & \url{https://huggingface.co/sentence-transformers/all-MiniLM-L6-v2} \\
Convergence entropy encoder & bert\textendash base\textendash uncased & \url{https://huggingface.co/bert-base-uncased} \\
Toxicity classifier & ToxiGen RoBERTa & \url{https://huggingface.co/tomh/toxigen_roberta} \\
\bottomrule
\end{tabularx}
\end{table}

\clearpage
\section{A day in the life of an agent}

This box illustrates how the abstract action rules in the simulator translate into a concrete sequence of platform interactions.

 \begin{tcolorbox}[colback=blue!5!white,colframe=blue!50!black,title={A day in the life},enhanced,sharp corners,left=6pt,right=6pt,top=4pt,bottom=4pt,float=htbp]
 \small
 When an agent is activated in a given round, a typical day might look like this:

 \begin{itemize}[nosep,leftmargin=*]
 \item 10:00 AM (Round 10). The agent is activated; according to their profile, they will perform two actions in this round.
 \item \textbf{Mentions:} Before those two actions, the agent first takes a “free” action and checks their recent mentions. If any are found, the agent writes a short reply in the thread.

 \item \textbf{Round action 1:} The simulator offers \texttt{[COMMENT, SHARE\_LINK, NONE]}; the LLM chooses \texttt{SHARE\_LINK}.
 \begin{itemize}[nosep,leftmargin=1em]
     \item Selects an article from the local news database matching interests; e.g., "New battery tech for grid storage."
     \item Reads the article and generates commentary, posted as a root submission with a URL to the source.
 \end{itemize}
 \item \textbf{Round action 2:} The simulator offers \texttt{[READ, POST, NONE]}; the LLM chooses \texttt{READ}.
 \begin{itemize}[nosep,leftmargin=1em]
     \item Reads a recommended post (root + comments up to the configured depth).
     \item Lurking behavior; no follow‑up action.
 \end{itemize}
 \item After two actions, the agent becomes inactive.
 \item 5:00 PM (Round 17). The agent is activated again; they will perform two actions.
 \item \textbf{Mentions:} The agent has a free reply action again.
 \item \textbf{Round action 1:} The simulator offers \texttt{[READ, COMMENT, NONE]}; the LLM chooses \texttt{COMMENT}.
 \begin{itemize}[nosep,leftmargin=1em]
 \item Retrieves candidate posts from the recommender and selects one uniformly at random.
 \item Fetches thread context via \texttt{/post\_thread} (last $K=\texttt{max\_length\_thread\_reading}$ items).
 \item Composes a concise, on‑topic reply.
 \item Prompted whether to like or dislike the post; reacts accordingly.
 \item The agent's interests are updated with topics associated with the root post.
 \end{itemize}
 \item \textbf{Round action 2:} The simulator offers \texttt{[SEARCH, COMMENT, NONE]}; the LLM chooses \texttt{NONE}. This is equivalent to observing the feed and doing nothing.
 \item The round budget is exhausted; the agent becomes inactive until sampled again.
 \end{itemize}
 \end{tcolorbox}


\newpage
\section{Core-periphery inference}



 The core–periphery structure is a canonical feature of social networks, extensively documented since the foundational work of Borgatti and Everett \cite{borgatti2000models} and formalized in modern detection frameworks \cite{rombach2014coreper}. In an SBM, nodes belong to latent groups (blocks) and edge probabilities depend only on the groups of the endpoints; the two‑block core–periphery variant encodes a dense core (high core–core and core–periphery probabilities) and a sparse periphery (low periphery–periphery probability), yielding a hub‑and‑spoke structure. This corresponds to the “two‑block” side of the clarified core–periphery typology \cite{gallagher2021clarif}, in contrast to layered k‑core decompositions.

\ Inference uses Gibbs updates (\(n_{\text{gibbs}}=100\)) and a Markov chain Monte Carlo (MCMC) length proportional to graph size (\(n_{\text{mcmc}}=10\,\lvert V\rvert\)), repeated over independent runs (\(n=5\)) for robustness. From each run we draw multiple posterior label sets in four 25‑sample windows, and also compute a 50‑sample consensus. For each sampled partition, we compute complementary quality criteria on the analyzed component: densities within the core and periphery and across the cut (core–periphery density), modularity of the two‑block partition (weighted), assortativity by the core/periphery label, and a minimum description length (MDL) score. Partitions are ranked by a composite score that balances within‑core density (0.3), core–periphery coupling (0.3), modularity (0.2), and a normalized MDL contribution (0.2); we also summarize variability in estimated core size across valid samples. 

 For characterization and visualization, we rank core members by degree and weighted degree and render the LCC with an edge‑weighted spring layout to highlight core concentration and core–periphery coupling. For comparisons, we construct an analogous Voat reply network for each matched 30‑day comparison window (\(n=30\)) using identical node/edge definitions; descriptors are computed per window and core–periphery inference is performed independently on each window’s LCC to ensure scope parity.

\newpage
\section{Embedding-threshold validation}


To validate the semantic similarity thresholds used in our analysis, we benchmarked the \texttt{all-MiniLM-L6-v2} sentence transformer model \cite{reimers-2019-sentence-bert} on the Semantic Textual Similarity Benchmark (STS-B) \cite{Cer2017}. The model achieved a Spearman correlation of $\rho = 0.82$ on the test set ($n = 1{,}379$ sentence pairs), consistent with published benchmarks \cite{reimers-2019-sentence-bert}. We then compared predicted cosine similarity with human-annotated STS-B scores. Pairs with predicted cosine similarity in the range $[0.5, 0.6)$ corresponded to a median gold score of $0.42$ (on a normalized 0--1 scale), indicating sentences that ``share some details but are not equivalent.'' Pairs in the $[0.6, 0.7)$ range yielded a median gold score of $0.60$, representing sentences that are ``roughly equivalent but differ in important information.'' These calibration results inform our downstream similarity thresholds. In the illustrative thread shown in the main manuscript's Validation section, pairs with minimal thematic overlap score 0.45--0.50 (below the ``share some details'' threshold), while pairs sharing vocabulary like ``creators'' and ``government'' reach 0.72--0.73, well above ``roughly equivalent.''

To visualize the shared embedding space, we render 2D t-distributed stochastic neighbor embedding (t-SNE) projections on a subset comprising all simulation items and a uniform sample of Voat points (up to 2{,}000). We use a cosine metric with perplexity 80. The Voat retrieval sets comprise the full v/technology corpus (2014–2020) for the corresponding content type, and the simulation uses all texts from the 30‑day run. The same encoder and preprocessing are applied across corpora.

\newpage
\section{Multi-run analysis and statistical inference}

A critical limitation of prior generative ABM studies is reliance on single simulation runs, which precludes uncertainty quantification and sensitivity assessment \cite{larooij2025do}. Multi-run replication is also consistent with established ABM validation practices: stochastic interaction can generate path dependence, so run-to-run variability is part of what must be characterized rather than averaged away \cite{fagiolo2007critical_validation,brown2005path_dependence_validation}. To address this gap and enable statistical comparison with matched Voat windows, we conducted 30 independent simulation runs with identical parameters but different random seeds (42--71). This replication design allows us to distinguish systematic patterns from stochastic variation and to compute confidence intervals for all reported metrics. Each run produces a complete 30-day trajectory stored in a SQLite database. We extract metrics from each run using a standardized pipeline. For the empirical benchmark, we use 30 matched, non-overlapping 30-day Voat comparison windows. The same 30 windows are used for activity, network, toxicity, and repeated-interaction comparisons. For the main 30-run simulation benchmark and the matched 30-window Voat comparison, we report 99\% confidence intervals based on the $t$-distribution. For the smaller sensitivity analyses, when interval estimates are reported, we use percentile bootstrap 99\% confidence intervals with 5{,}000 resamples and seed 42.

\newpage
\section{Voat Benchmark Statistics}

Table~\ref{tab:voat-benchmark} reports the aggregate benchmark statistics for the 30 matched, non-overlapping 30-day Voat windows introduced above.

\begin{table}[htbp]
\centering
\caption{Voat v/technology benchmark statistics across 30 matched, non-overlapping 30-day comparison windows. Intervals are 99\% $t$-distribution confidence intervals.}
\label{tab:voat-benchmark}
\small
\begin{tabular}{lrrr}
\toprule
Metric & Mean & Std & 99\% CI \\
\midrule
Root posts (threads) & 568.7 & 146.1 & [495.2, 642.2] \\
Comments & 733.1 & 262.8 & [600.9, 865.4] \\
Unique users & 591.1 & 170.4 & [505.3, 676.8] \\
Avg daily active users & 32.7 & 9.8 & [27.7, 37.6] \\
Avg thread size & 2.25 & 0.22 & [2.14, 2.36] \\
Mean toxicity & 0.119 & 0.020 & [0.108, 0.129] \\
\botrule
\end{tabular}
\end{table}

\newpage
\section{Confidence Intervals and Simulation Statistics}


For the main 30-run simulation benchmark and the matched 30-window Voat comparison, confidence intervals for simulation and Voat metrics are computed with the Student $t$ distribution:
\begin{equation}
\text{CI}_{99\%} = \bar{x} \pm t_{0.995,\,n-1}\frac{s}{\sqrt{n}},
\end{equation}
where $\bar{x}$ is the sample mean, $s$ is the sample standard deviation, and $n=30$ for both the simulation runs and the matched Voat windows, giving $t_{0.995,29}\approx2.756$. We report the raw $t$-based intervals; for bounded proportions, rounding can produce limits slightly outside $[0,1]$. For the smaller sensitivity analyses below, when interval estimates are reported, we instead use percentile bootstrap 99\% confidence intervals with 5{,}000 resamples and seed 42.

For the two tail-shape claims used in the main text (root posts per user and network degree), we supplement these interval summaries with a separate discrete tail analysis following Clauset, Shalizi, and Newman \cite{clauset2009powerlaw}. Each 30-day simulation run and each matched 30-day Voat window is fit separately rather than pooled. For each sample, we estimate $x_{\min}$ by minimizing the Kolmogorov--Smirnov distance, estimate the power-law exponent $\alpha$ by maximum likelihood above $x_{\min}$, and compute a bootstrap goodness-of-fit $p$-value from 1{,}000 synthetic samples that preserve the empirical body below $x_{\min}$ while drawing the upper tail from the fitted discrete power law. We treat a window as power-law-compatible when $n_{\text{tail}}\ge 50$ and $p\ge 0.10$. On the same support, we compare the power law against lognormal, truncated power-law, and exponential alternatives using normalized log-likelihood ratio tests.


Table~\ref{tab:sim-full} presents the corresponding simulation metrics with 99\% $t$-distribution confidence intervals.

\begin{table}[htbp]
\centering
\caption{Simulation metrics across 30 independent runs (99\% $t$-distribution CI).}
\label{tab:sim-full}
\small
\begin{tabular}{llrrr}
\hline
\multicolumn{2}{c}{Metric}                                           & Mean     & Std    & 99\% CI                  \\ \hline
\multirow{7}{*}{\textit{Activity Metrics}} & Total items             & 1496.9   & 68.1   & {[}1462.7, 1531.2{]}     \\
                                           & Comments                & 903.6    & 46.4   & {[}880.2, 926.9{]}       \\
                                           & Root posts (threads)    & 593.4    & 32.1   & {[}577.2, 609.5{]}       \\
                                           & Unique users            & 610.3    & 29.9   & {[}595.2, 625.3{]}       \\
                                           & Mean posts/user         & 2.46     & 0.10   & {[}2.40, 2.51{]}         \\
                                           & Avg thread length       & 2.53     & 0.08   & {[}2.48, 2.57{]}         \\
                                           & Mean daily active users & 37.3     & 1.43   & {[}36.6, 38.0{]}         \\ \hline
\multirow{8}{*}{\textit{Network Metrics}}  & Nodes (full network)    & 610.3    & 29.9   & {[}595.2, 625.3{]}       \\
                                           & Edges                   & 830.5    & 41.4   & {[}809.7, 851.4{]}       \\
                                           & Avg degree              & 2.72     & 0.12   & {[}2.66, 2.78{]}         \\
                                           & Avg clustering          & 0.017    & 0.005  & {[}0.015, 0.020{]}       \\
                                           & Density                 & 0.0102   & 0.0008 & {[}0.0098, 0.0106{]}     \\
                                           & LCC nodes               & 405.6    & 20.2   & {[}395.4, 415.8{]}       \\
                                           & LCC ratio               & 0.665    & 0.016  & {[}0.657, 0.673{]}       \\
                                           & Modularity              & $-0.121$ & 0.016  & {[}$-0.129$, $-0.113${]} \\ \hline
\multirow{3}{*}{\textit{Core-Periphery}}   & Core nodes              & 79.6     & 9.2    & {[}74.9, 84.2{]}         \\
                                           & Periphery nodes         & 326.0    & 18.3   & {[}316.8, 335.2{]}       \\
                                           & Core \% of LCC          & 19.6\%   & 2.1\%  & {[}18.6\%, 20.7\%{]}     \\ \hline
\multirow{9}{*}{\textit{Toxicity}}         & Mean toxicity (overall) & 0.143    & 0.016  & {[}0.135, 0.151{]}       \\
                                           & Median toxicity         & 0.009    & 0.002  & {[}0.009, 0.010{]}       \\
                                           & Toxicity std dev        & 0.274    & 0.016  & {[}0.266, 0.282{]}       \\
                                           & P90 toxicity            & 0.654    & 0.098  & {[}0.604, 0.703{]}       \\
                                           & P95 toxicity            & 0.893    & 0.030  & {[}0.877, 0.908{]}       \\
                                           & Frac $>0.5$             & 12.6\%   & 1.7\%  & {[}11.8\%, 13.5\%{]}     \\
                                           & Frac $>0.8$             & 7.7\%    & 1.3\%  & {[}7.0\%, 8.3\%{]}       \\
                                           & Comment mean toxicity   & 0.161    & 0.019  & {[}0.151, 0.170{]}       \\
                                           & Post/news mean toxicity & 0.116    & 0.017  & {[}0.107, 0.124{]}       \\ \hline
\end{tabular}
\end{table}

\clearpage
\newpage
\section{Additional Network Results}

\subsection{Weighted Degree Distribution}

Figure~\ref{fig:weighted-degree} shows the weighted degree distribution across all 60 networks in the main benchmark (30 simulation runs + 30 matched 30-day Voat windows). Weighted degree is node strength, computed as the sum of repeated-interaction edge weights incident on each node.

\begin{figure}[htbp]
\centering
\includegraphics[width=\textwidth]{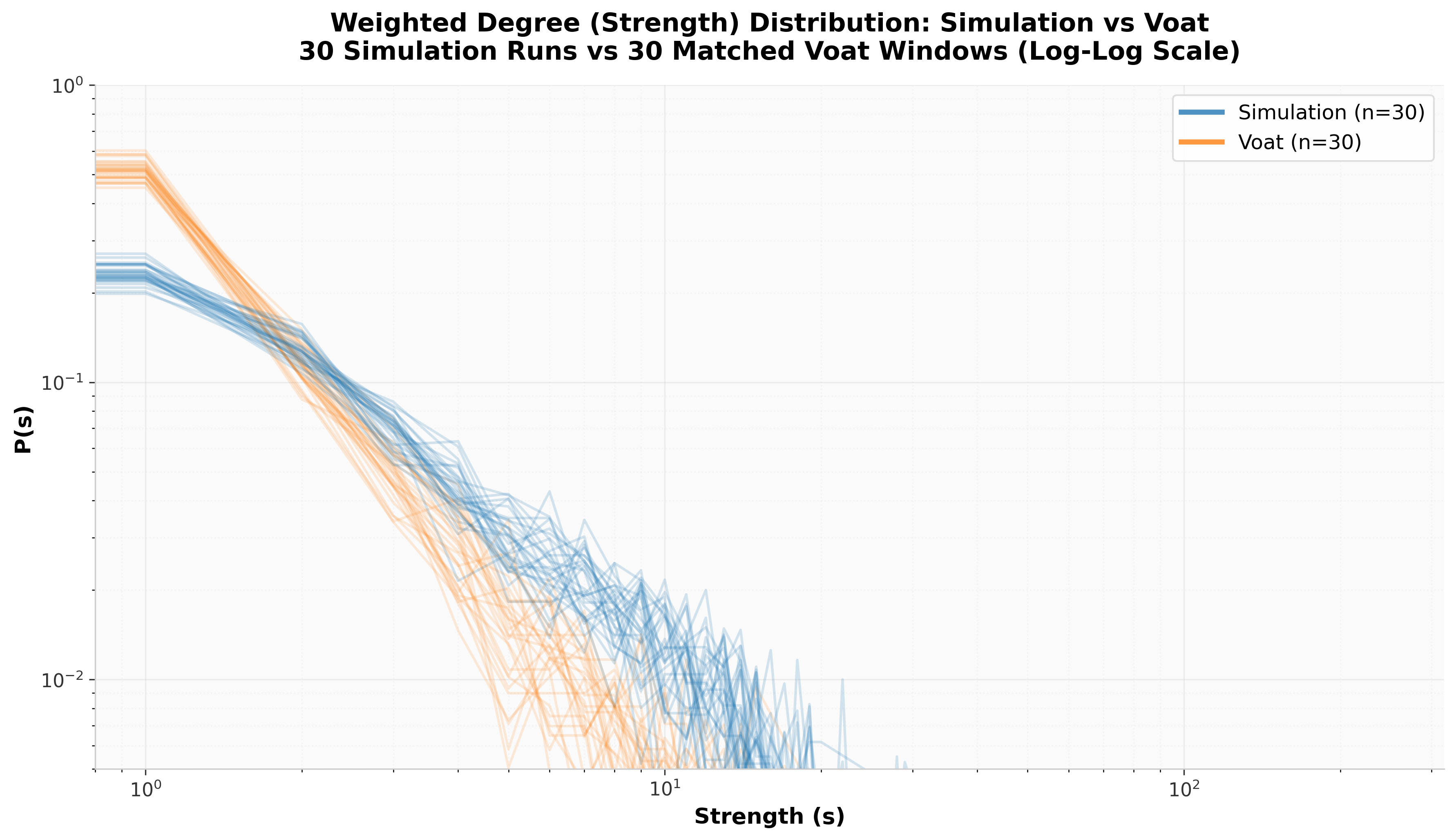}
\caption{Weighted degree (node strength) distributions for the same 60 benchmark networks used in the main paper: 30 simulation runs (blue) and 30 matched 30-day Voat comparison windows (orange). Each line corresponds to one network, plotted on log-log axes.}
\label{fig:weighted-degree}
\end{figure}

\newpage
\subsection{Repeated Interactions}

Figure~\ref{fig:repeated-interactions} analyzes repeated interactions in the same 60-network benchmark comparison. Here an edge counts as repeated when the same user pair interacts more than once within a 30-day window.

\begin{figure}[htbp]
\centering
\includegraphics[width=\textwidth]{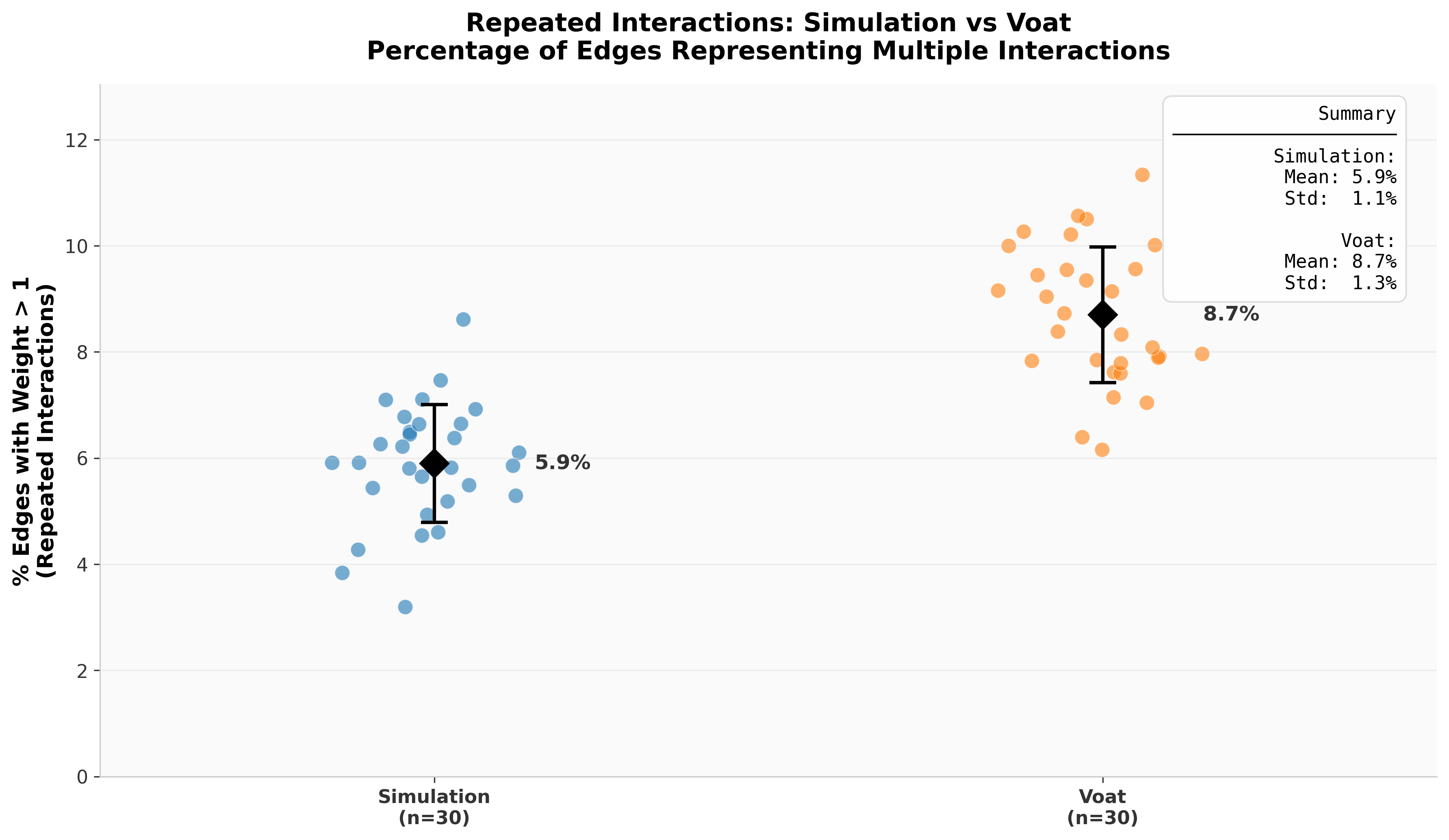}
\caption{Repeated-interaction rates in 30 simulation runs versus the 30 matched 30-day Voat comparison windows. Points are individual networks; black diamonds and bars show mean $\pm$ SD for the percentage of edges with weight $>1$.}
\label{fig:repeated-interactions}
\end{figure}

\clearpage
\subsection{Degree Distribution Analysis}

Figure~\ref{fig:degree-analysis} shows distribution-free hub-concentration diagnostics for the same 30 simulation runs and 30 matched Voat windows used in the main benchmark. Formal power-law fits are reported separately in Table~\ref{tab:tail-shape}.

\begin{figure}[htbp]
\centering
\includegraphics[width=\textwidth]{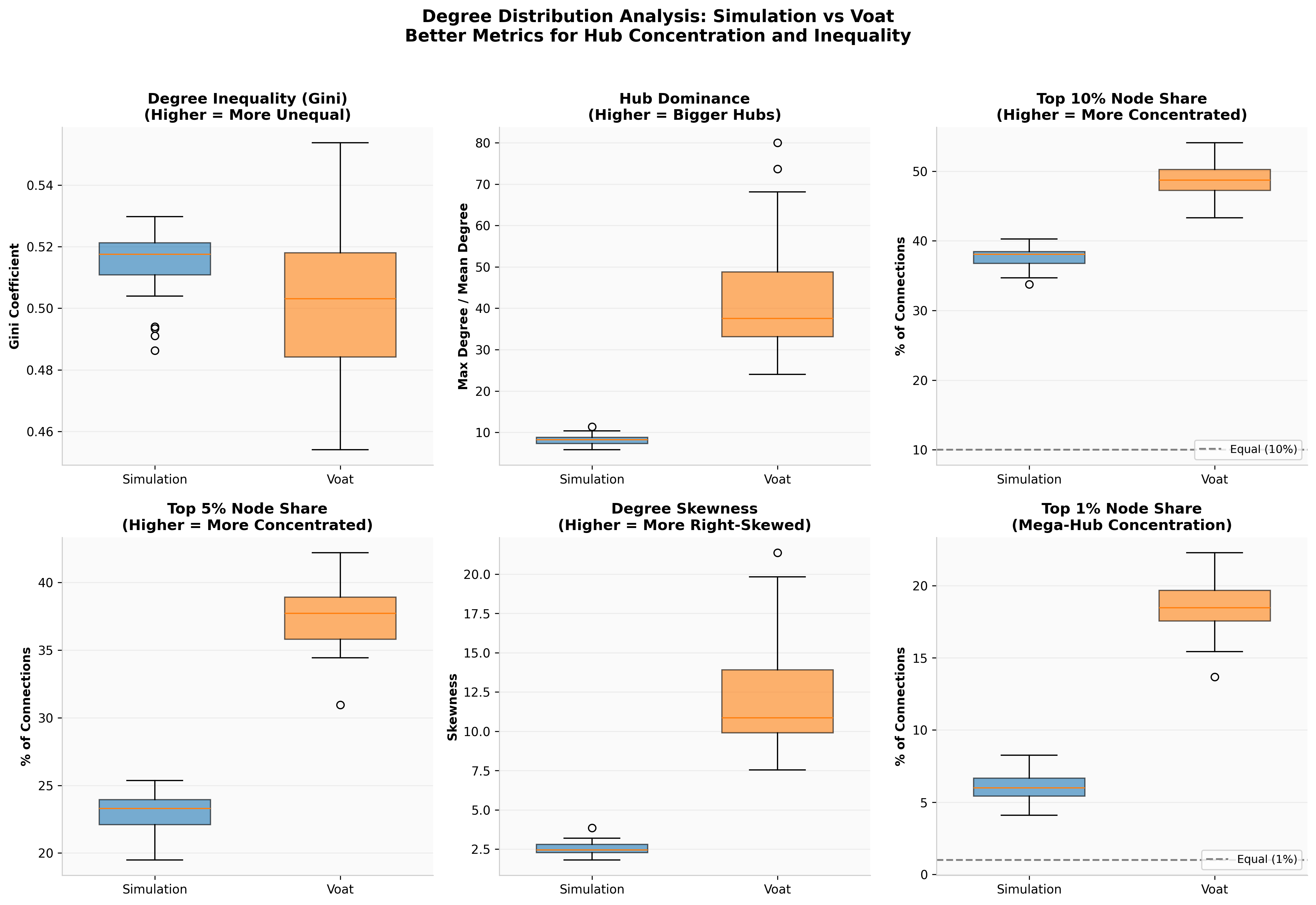}
\caption{Degree concentration diagnostics for simulation and matched Voat networks. The six panels compare Gini inequality, max/mean degree ratio, top-10\% share, top-5\% share, skewness, and top-1\% share across the 60 benchmark networks.}
\label{fig:degree-analysis}
\end{figure}

\subsection{Tail-Shape Validation}

Table~\ref{tab:tail-shape} summarizes the formal tail fits for the two heavy-tail claims used in the main text: root posts per user and positive network degree.

\begin{table}[htbp]
\centering
\caption{Compact tail-shape summary for the two main-text observables. Each 30-day run/window is fit separately. ``PL-compatible'' counts windows with $n_{\text{tail}}\ge 50$ and bootstrap goodness-of-fit $p\ge 0.10$ under the fitted discrete power law. ``Alt.'' reports the alternative family that most often outperforms the pure power law in a likelihood-ratio comparison. ``Top 1\% share'' is the median percentage of total mass held by the top 1\% of users or nodes.}
\label{tab:tail-shape}
\footnotesize
\begin{tabularx}{\textwidth}{@{} l l c c >{\raggedright\arraybackslash}p{0.28\textwidth} c @{}}
\toprule
\textbf{Observable} & \textbf{Corpus} & \textbf{PL-comp.} & \textbf{Med.\ $\alpha$} & \textbf{Alt.} & \textbf{Top 1\% share} \\
\midrule
Posts/user & Simulation & 27/30 & 2.88 & TPL preferred in 30/30 & 5.1\% \\
Posts/user & Voat & 9/30 & 2.06 & No dominant alternative; PL $>$ Exp in 19/30 & 30.4\% \\
Degree & Simulation & 5/30 & 2.83 & LN preferred in 26/30; TPL in 29/30 & 7.2\% \\
Degree & Voat & 28/30 & 2.18 & PL usually adequate; TPL in 2/30 & 20.3\% \\
\botrule
\end{tabularx}
\end{table}

The participation tail is heavy-tailed in both corpora, but not in the same way. Simulation runs usually admit a power-law fit only above $x_{\min}=2$, yet the truncated power law outperforms the pure power law in all 30 runs, consistent with the finite action-budget ceiling. Voat posting activity is far more concentrated by distribution-free metrics (median top-1\% share 30.4\% versus 5.1\%), but only 9 of 30 windows support a pure power-law fit, indicating a heavier and more heterogeneous winner-take-most participation pattern across windows rather than one stable family across the full benchmark.

The network result is sharper. Positive degree is power-law-compatible in 28 of 30 Voat windows, with median $x_{\min}=2$ and median $\alpha=2.18$, but only 5 of 30 simulation runs. In the simulation, lognormal and truncated alternatives beat the pure power law in 26/30 and 29/30 runs, respectively, and the fitted tail begins much farther out ($x_{\min}=6$). Distribution-free concentration metrics point in the same direction: the top 1\% of Voat nodes hold a median 20.3\% of total degree, compared with 7.2\% in simulation. This formalizes the main-text claim that the simulation underproduces extreme hubs even when average degree remains close to the empirical benchmark.

\newpage
\section{Toxicity Analysis Results}

\subsection{Toxicity Benchmark}

For the current 30-window benchmark, mean toxicity is higher in simulation than in Voat overall: 0.143 [0.135, 0.151] versus 0.119 [0.108, 0.129]. The mismatch is layer-specific. Simulated root posts/news items are much more toxic than Voat root submissions (0.116 [0.107, 0.124] vs.\ 0.0307 [0.0265, 0.0348]), whereas simulated comments are less toxic than Voat comments (0.161 [0.151, 0.170] vs.\ 0.189 [0.173, 0.205]). Thus the simulation preserves the direction of the posts-to-comments gradient but misallocates toxicity toward root-level content. The OAT sensitivity panel below should be read as a robustness probe around this mismatch, not as a replacement benchmark for the 30-day simulation--Voat comparison reported in the main text.

\subsection{Reply-Depth Mechanism Check}

We tested whether toxicity increases with reply depth beyond the coarse root/comment distinction. The observed Spearman association between depth and toxicity is $\rho = 0.122$. Under the depth-shuffle permutation null, the 95\% interval is [0.121, 0.127], with $p = 0.896$. The observed association therefore falls inside the null distribution. The apparent depth pattern is explained by the post/comment split rather than by within-thread escalation, consistent with the stateless, per-action design.

\newpage
\section{Sensitivity and Robustness Analyses}
\label{sec:si-sensitivity}

The one-at-a-time (OAT) sensitivity panel probes local calibration around the baseline specification rather than re-running the full 30-day benchmark. Each variant uses a separate 10-day setup with the same analysis pipeline as the main paper, and all deltas in this section are reported relative to the corresponding 10-day baseline condition c0. The baseline c0 retains the main-paper temperature setting of 0.6, so the temperature family compares that baseline against lower (0.3) and higher (0.9) variants. When interval estimates are used here, they are percentile bootstrap 99\% confidence intervals with 5{,}000 resamples and seed 42. We mark a change as material when the variant and baseline intervals do not overlap; this rule is encoded by the starred entries in Table~\ref{tab:sensitivity-matrix} and the outlined cells in Figure~\ref{fig:sensitivity-overview}. The main 30-run / 30-window benchmark remains the reference for validation; the OAT panel is a robustness probe for mechanism sensitivity around that benchmark.

\begin{table}[htbp]
\centering
\caption{Family-level summary of the 10-day OAT sensitivity panel. ``Material shifts'' denote variants whose 99\% bootstrap confidence intervals do not overlap the c0 baseline interval for at least one benchmark metric.}
\label{tab:sensitivity-suite-overview}
\small
\begin{tabularx}{\textwidth}{@{} >{\raggedright\arraybackslash}p{0.17\textwidth} >{\raggedright\arraybackslash}p{0.18\textwidth} >{\raggedright\arraybackslash}X >{\raggedright\arraybackslash}X @{}}
\toprule
\textbf{Family} & \textbf{Variants} & \textbf{Material shifts vs.\ c0} & \textbf{Interpretation} \\
\midrule
Persona cues & Neutral persona (c1), no-politics persona (c2) & c1 reduces mean toxicity by 46.7\%, toxicity p90 by 10.7\%, frac$>0.5$ by 51.5\%, and comment toxicity by 26.2\%; c2 overlaps baseline on all tracked metrics & Toxicity is strongly sensitive to persona framing, while activity, topic coverage, and entropy remain close to baseline \\
Temperature & Low temperature 0.3 (c3), high temperature 0.9 (c4) & Only posts/user at c3 shows a non-overlapping interval (+15.5\%); other changes overlap baseline & Sampling temperature has weak local effects relative to other knobs \\
Budget slope & Flat Zipf slope 1.5 (c5), steep slope 3.5 (c6) & c5 raises posts/day by 80.1\%, comments/day by 94.5\%, DAU by 52.2\%, and number of topics by 75.2\%, while lowering core share by 51.3\%; c6 mainly lowers entropy by 3.9--4.5\% & Flatter budgets broaden participation and topical breadth; steeper budgets chiefly compress local entropy with limited activity change \\
Comment-to-post ratio & CPR 2:1 (c7), CPR 50:1 (c8) & c7 lowers comments/day by 26.7\% and thread length by 22.3\% while raising density by 66.7\%; c8 raises posts/day by 58.4\%, comments/day by 128.4\%, DAU by 42.3\%, and number of topics by 49.6\%, while lowering density by 62.6\% and core share by 59.4\% & Reply availability is a major structural lever, especially for activity volume and network concentration \\
Churn & Low churn (c9), high churn (c10) & c10 raises posts/day by 42.7\%, comments/day by 47.5\%, DAU by 40.0\%, and number of topics by 38.0\%; c9 overlaps baseline on starred metrics & Turnover primarily changes activity scale and topic breadth, with limited effect on toxicity \\
\botrule
\end{tabularx}
\end{table}

\begin{figure}[htbp]
\centering
\includegraphics[width=\textwidth]{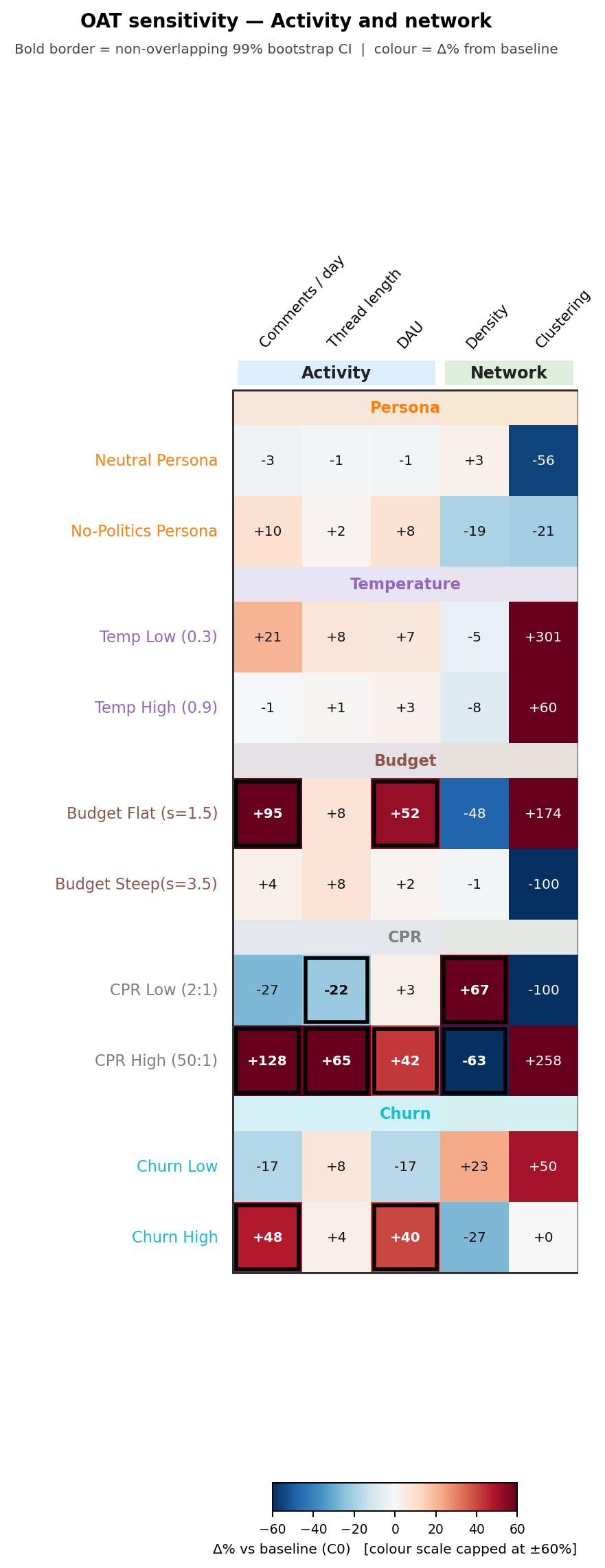}
\vspace{0.5em}
\includegraphics[width=\textwidth]{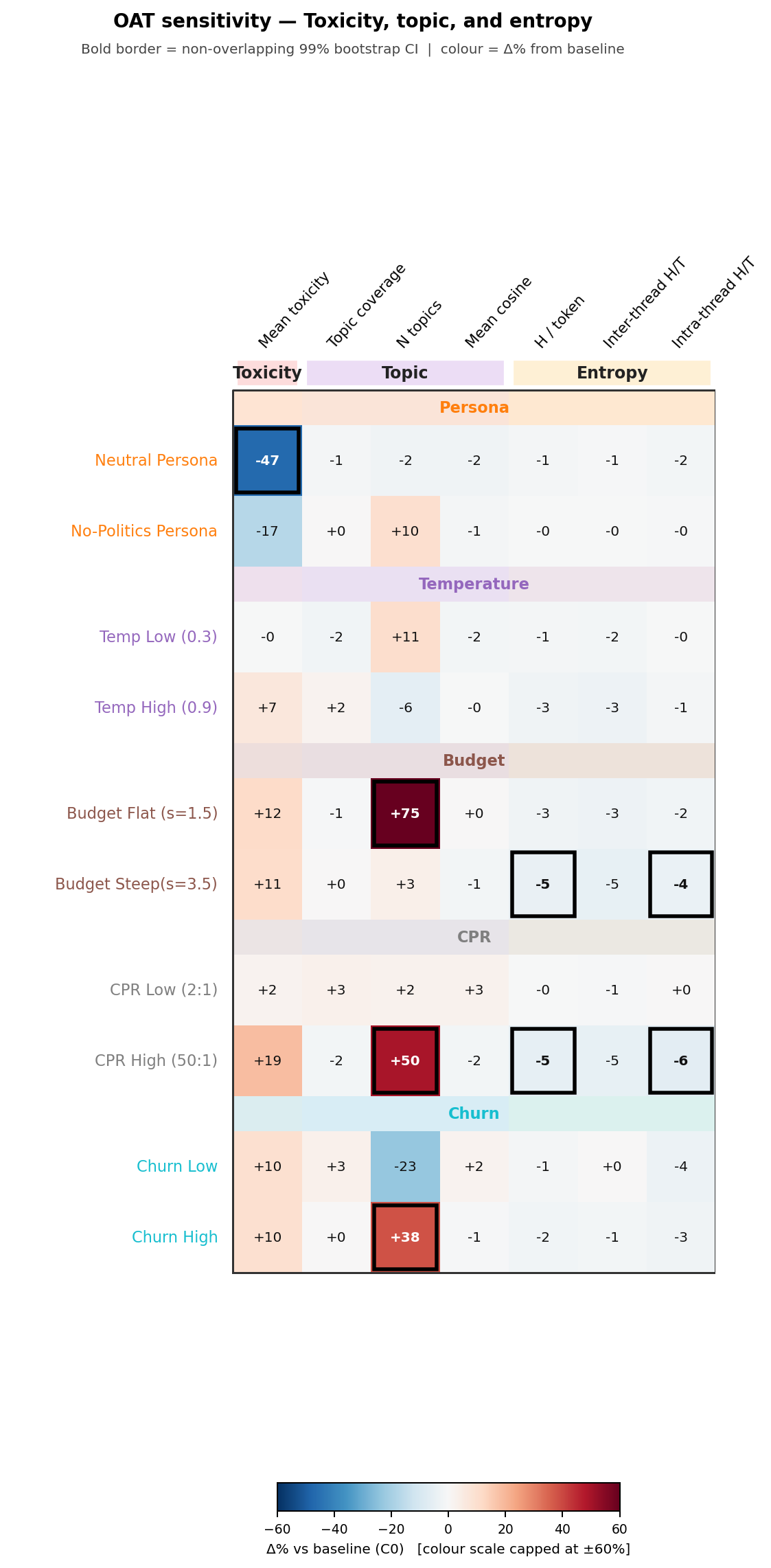}
\caption{Overview of the 10-day OAT sensitivity panel relative to baseline c0. Top: activity and network metrics. Bottom: toxicity, topic, and entropy metrics. Cell values show percent change from baseline; outlined cells mark variants whose 99\% bootstrap confidence intervals do not overlap the c0 interval. These panels summarize local robustness around the baseline specification and are not directly comparable in scale to the 30-day benchmark tables.}
\label{fig:sensitivity-overview}
\end{figure}

\begin{sidewaystable}[htbp]
\centering
\caption{Full 10-day OAT sensitivity matrix. Baseline c0 values are means for the 10-day baseline condition. Each cell gives percent change relative to c0; `*' marks non-overlapping 99\% bootstrap confidence intervals versus c0. Column abbreviations: Ntrl = neutral persona; NoPol = no-politics persona; T0.3/T0.9 = temperature 0.3/0.9; B1.5/B3.5 = budget Zipf slope 1.5/3.5; CPR2/CPR50 = comment-to-post ratio 2:1 / 50:1; ChL/ChH = low / high churn.}
\label{tab:sensitivity-matrix}
\tiny
\setlength{\tabcolsep}{2pt}
\begin{tabular}{l r r r r r r r r r r r}
\toprule
\textbf{Metric} & \textbf{C0} & \textbf{Ntrl} & \textbf{NoPol} & \textbf{T0.3} & \textbf{T0.9} & \textbf{B1.5} & \textbf{B3.5} & \textbf{CPR2} & \textbf{CPR50} & \textbf{ChL} & \textbf{ChH} \\
\midrule
\multicolumn{12}{l}{\textit{Activity}} \\
Posts / day & 5.36 & -1.9 & +8.6 & +12.3 & -0.3 & +80.1* & -2.9 & +7.6 & +58.4* & -17.4 & +42.7* \\
Comments / day & 2.47 & -2.6 & +9.8 & +20.8 & -0.8 & +94.5* & +3.7 & -26.7* & +128.4* & -17.4 & +47.5* \\
Posts / user & 1.48 & +3.8 & +4.3 & +15.5* & +0.8 & +37.5* & +0.3 & +7.0 & +22.0* & +7.0 & +9.5 \\
Avg thread length & 1.88 & -1.1 & +1.9 & +7.6 & +0.7 & +8.2 & +8.0 & -22.3* & +64.6* & +7.7 & +4.1 \\
Daily active users & 4.66 & -1.5 & +8.5 & +6.9 & +2.6 & +52.2* & +1.7 & +3.4 & +42.3* & -16.9 & +40.0* \\
\midrule
\multicolumn{12}{l}{\textit{Network}} \\
Density & 0.261 & +3.0 & -19.1 & -5.1 & -7.7 & -47.9 & -1.5 & +66.7* & -62.6* & +22.8 & -27.2 \\
Avg clustering & 0.0046 & -55.6 & -20.8 & +300.8 & +60.3 & +174.0 & -100.0 & -100.0 & +258.0 & +50.0 & +0.2 \\
Core \% & 24.22 & +23.4 & -11.4 & -13.8 & +11.1 & -51.3* & +26.1 & +75.2 & -59.4* & +27.1 & -1.4 \\
Core density & 0.800 & +0.0 & -16.7 & -16.7 & -16.7 & -4.2 & +4.2 & -8.3 & -20.8 & -25.1 & -14.2 \\
Modularity & -0.072 & -75.2 & -96.7 & -130.6 & -81.9 & +94.9 & -113.6 & -172.1 & +30.5 & -58.2 & -71.6 \\
\midrule
\multicolumn{12}{l}{\textit{Toxicity}} \\
Mean toxicity & 0.459 & -46.7* & -17.3 & -0.3 & +6.9 & +11.6 & +11.1 & +2.0 & +18.7 & +9.8 & +9.8 \\
Toxicity p90 & 0.932 & -10.7* & -2.9 & +1.2 & +2.5 & +2.8 & +2.0 & +0.9 & +2.5 & +1.8 & +2.0 \\
Frac $>0.5$ & 0.491 & -51.5* & -23.8 & -3.5 & +4.8 & +8.1 & +8.9 & -0.1 & +18.5 & +7.9 & +9.8 \\
Comment tox.\ mean & 0.603 & -26.2* & -2.6 & +3.9 & +1.0 & +6.6 & +8.6 & +9.8 & -0.1 & +0.4 & +5.8 \\
\midrule
\multicolumn{12}{l}{\textit{Topic}} \\
Topic coverage & 0.969 & -1.0 & +0.1 & -2.3 & +2.3 & -0.5 & +0.5 & +3.2 & -1.6 & +3.2 & +0.3 \\
Mean cosine sim. & 0.630 & -2.4 & -1.4 & -1.8 & -0.1 & +0.3 & -1.4 & +2.8 & -1.8 & +2.0 & -0.8 \\
N topics & 12.10 & -2.5 & +9.9 & +10.7 & -5.8 & +75.2* & +3.3 & +2.5 & +49.6* & -23.1 & +38.0* \\
\midrule
\multicolumn{12}{l}{\textit{Entropy}} \\
H / token & 0.272 & -1.2 & -0.3 & -1.1 & -2.5 & -2.8 & -4.5* & -0.4 & -5.5* & -1.3 & -1.9 \\
Inter-thread H/T & 0.267 & -0.9 & -0.2 & -1.6 & -3.2 & -3.0 & -4.8 & -0.5 & -5.0 & +0.1 & -1.5 \\
Intra-thread H/T & 0.281 & -1.8 & -0.5 & -0.1 & -1.1 & -2.3 & -3.9* & +0.0 & -6.2* & -3.6 & -2.5 \\
\botrule
\end{tabular}
\end{sidewaystable}

The quantitative pattern is more selective than the old single-variant summary suggested. Persona manipulation remains the cleanest toxicity lever: neutralizing persona cues nearly halves mean toxicity and the share of clearly toxic items, yet leaves comments/day, topic coverage, and entropy near baseline. The no-politics variant, by contrast, moves the same metrics only modestly and keeps overlapping intervals throughout. This reinforces the main-text interpretation that toxicity is sensitive to persona framing rather than being hard-wired by the evaluation pipeline alone.

Temperature is comparatively weak. Lower temperature raises posts/user by 15.5\% with a non-overlapping interval, but most other activity, toxicity, topic, and entropy metrics remain within the bootstrap-overlap band. Budget slope, by contrast, is consequential: the flatter budget profile (c5) broadly inflates participation and topic breadth while shrinking the inferred core share, whereas the steeper profile (c6) leaves activity closer to baseline but lowers entropy modestly. The OAT panel therefore suggests that participation-allocation assumptions matter more than local sampling stochasticity for matching community-scale behavior.

The comment-to-post ratio and churn controls are the strongest structural levers. CPR50 (c8) simultaneously raises posts/day, comments/day, thread length, daily active users, and topic count while making the network sparser and the core smaller; CPR2 (c7) pushes in the opposite direction for comments/day and thread length while increasing density. High churn (c10) expands activity and topical breadth without strongly moving toxicity, whereas low churn (c9) mostly overlaps baseline. Taken together, these results indicate that the current mismatch profile is not driven by one universal instability: toxicity is primarily persona-sensitive, while activity and network concentration are most responsive to budget, reply availability, and turnover.

\newpage
\section{Embedding Similarity Results}

\subsection{Alternative Visualizations}

Figure~\ref{fig:umap} shows a UMAP projection for the same selected comment run used in the main-text embedding figure: run11, chosen because it has the highest mean comment-to-comment cosine across the 30 simulation runs.

\begin{figure}[htbp]
\centering
\includegraphics[width=\textwidth]{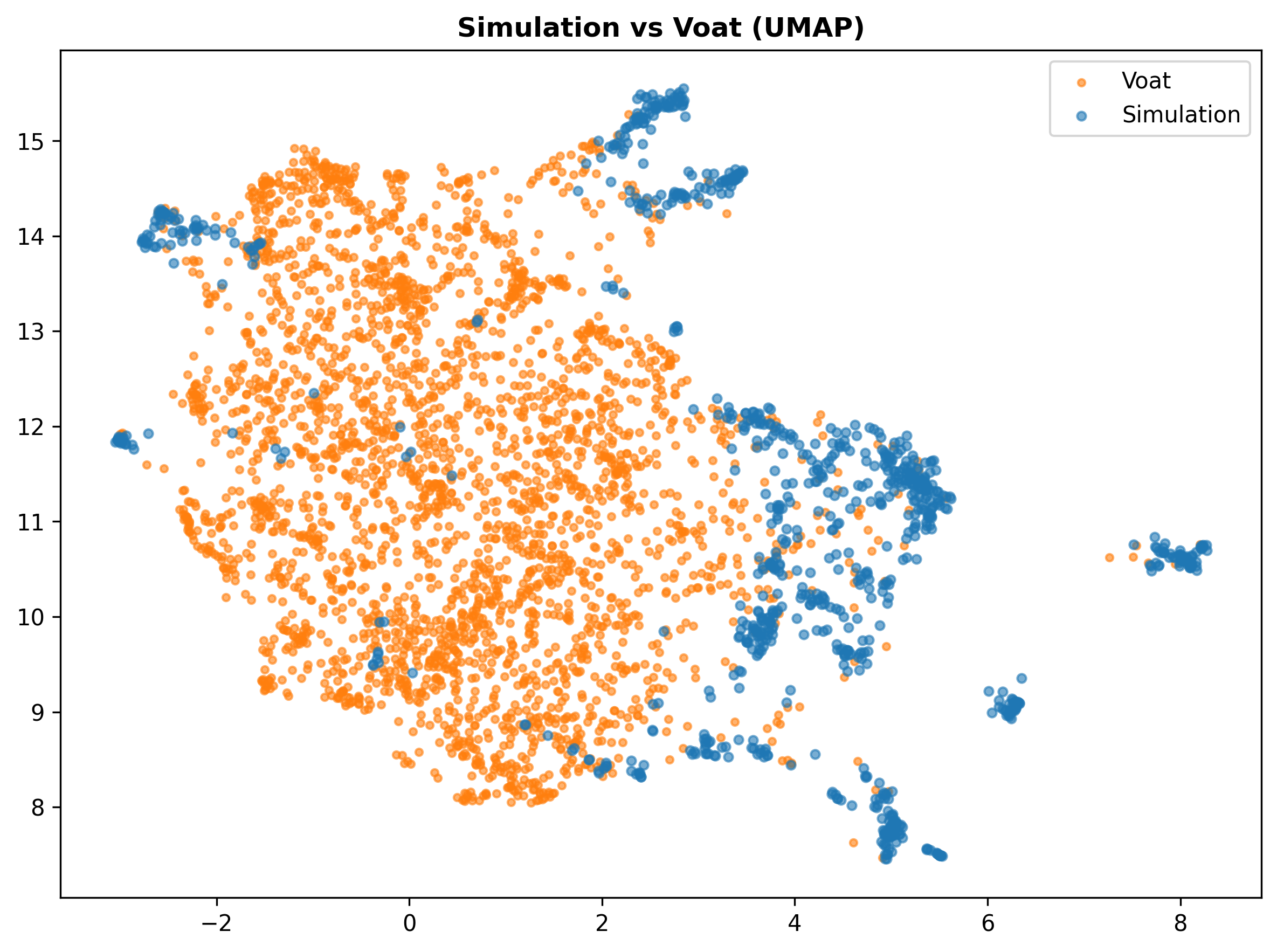}
\caption{UMAP projection of comment embeddings for simulation run11 (blue) against a uniform sample of comments from the full Voat v/technology corpus (orange). The run was selected because it has the highest mean comment cosine in the 30-run benchmark.}
\label{fig:umap}
\end{figure}

\newpage
\subsection{Combined UMAP and t-SNE}

Figure~\ref{fig:combined-embed} places UMAP and t-SNE side by side for the same comment-level embedding set, allowing the two reductions to be compared directly.

\begin{figure}[htbp]
\centering
\includegraphics[width=\textwidth]{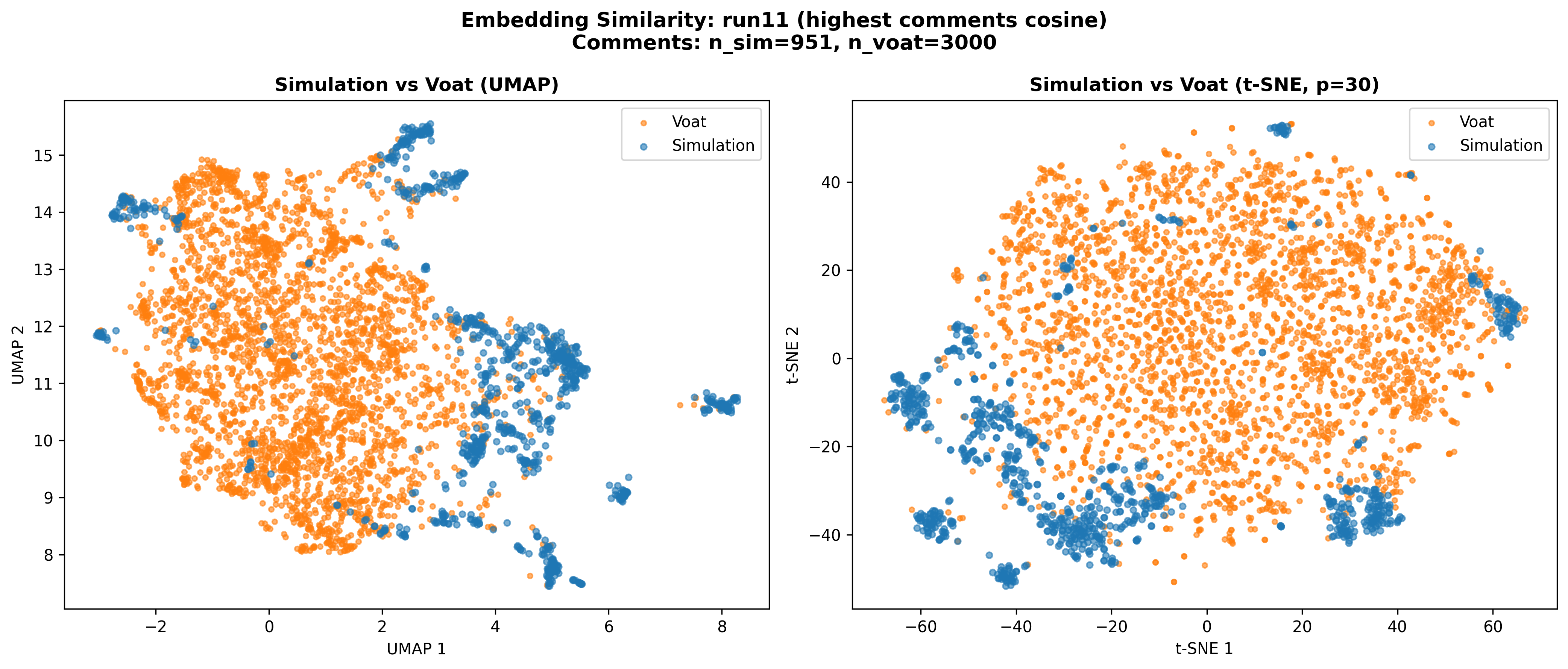}
\caption{Combined UMAP (left) and t-SNE (right) projections for comment embeddings from the selected simulation run11 against a sampled Voat comment background. Both panels use the same all-MiniLM-L6-v2 embedding space as the main-text nearest-neighbor analysis.}
\label{fig:combined-embed}
\end{figure}

\subsection{Embedding Similarity Statistics}

Table~\ref{tab:embedding-full} reports the aggregate nearest-neighbor similarity metrics.

\begin{table}[htbp]
\centering
\caption{Embedding similarity metrics (simulation to Voat) with 99\% $t$-distribution CI across 30 simulation runs.}
\label{tab:embedding-full}
\begin{tabular}{lrrr}
\toprule
Metric & Mean & Std & 99\% CI \\
\midrule
Comments mean cosine & 0.573 & 0.013 & [0.566, 0.580] \\
Comments median cosine & 0.568 & 0.015 & [0.560, 0.576] \\
Posts mean cosine & 0.607 & 0.004 & [0.604, 0.609] \\
Posts median cosine & 0.612 & 0.005 & [0.609, 0.614] \\
\botrule
\end{tabular}
\end{table}

\newpage
\section{Topic Modeling Details}

This section first reports aggregate topic-matching statistics and then provides the complete thread-level topic list and top Voat matches for the representative run01 referenced by the main-text topic-match table.

\subsection{Full Topic Statistics}

Table~\ref{tab:topic-full} presents the complete topic-matching statistics.

\begin{table}[htbp]
\centering
\caption{Topic modeling comparison metrics (99\% $t$-distribution CI across 30 simulation runs).}
\label{tab:topic-full}
\begin{tabular}{lrrr}
\toprule
Metric & Mean & Std & 99\% CI \\
\midrule
\multicolumn{4}{l}{\textit{Document-Level}} \\
Topic coverage & 90.0\% & 5.6\% & [87.2\%, 92.9\%] \\
Mean cosine similarity & 0.614 & 0.010 & [0.609, 0.619] \\
Median cosine similarity & 0.613 & 0.012 & [0.607, 0.619] \\
\midrule
\multicolumn{4}{l}{\textit{Thread-Level}} \\
Topic coverage & 99.6\% & 1.2\% & [99.0\%, 100.2\%] \\
Mean cosine similarity & 0.687 & 0.014 & [0.680, 0.693] \\
\botrule
\end{tabular}
\end{table}

\newpage
\subsection{Full Topic Match List (Representative Run)}

To improve transparency and reduce dependence on cherry-picked examples, Table~\ref{tab:topic-matches-run01} lists the complete set of thread-level simulation topics for the representative run (run01) together with their top-1 Voat match (highest cosine similarity) under the same centroid-based matching procedure used in the main text (threshold 0.50). Topic labels are the BERTopic top-word labels (lightly cleaned for readability).

\begin{table}[htbp]
\centering
\caption{Full thread-level topic match list for the representative run (run01). Topic labels are automatically generated by BERTopic from the top representative words for each cluster. For each simulation topic, we report its highest-similarity Voat topic (cosine similarity between topic centroids, threshold 0.50) and the corresponding topic sizes (number of threads assigned).}
\label{tab:topic-matches-run01}
\scriptsize
\begin{tabular}{r p{0.31\textwidth} r p{0.31\textwidth} r c}
\toprule
\textbf{Sim topic} & \textbf{Simulation label} & \textbf{N} & \textbf{Closest Voat label} & \textbf{N} & \textbf{Cos} \\
\midrule
0 & media just news like & 96 & trump jews jewish goldman sachs & 41 & 0.73 \\
1 & wages trial companies collusion & 34 & valley silicon valley silicon layoffs & 35 & 0.68 \\
2 & bitchute platforms just platform & 33 & copyright happy happy tpp happy & 27 & 0.74 \\
3 & businesses maybe small just & 31 & data privacy breaches track & 40 & 0.82 \\
4 & tech big tech regulation development & 29 & data privacy breaches track & 40 & 0.68 \\
5 & energy solar evs electric & 28 & tesla solar electric energy & 307 & 0.61 \\
6 & apple email tech proton & 25 & google com search results & 259 & 0.75 \\
7 & opal white people diversity & 22 & trump jews jewish goldman sachs & 41 & 0.70 \\
8 & intel pcs gaming gaming pcs & 21 & intel amd security spectre & 56 & 0.71 \\
9 & browser web activex download & 18 & firefox browser mozilla opera & 67 & 0.71 \\
10 & download private just trackers & 18 & internet china open global & 34 & 0.67 \\
11 & companies just intel repair & 18 & iot internet things internet secure internet & 21 & 0.76 \\
12 & open source source open linux & 17 & open source source open osg & 26 & 0.70 \\
13 & medicine genetic science pharma & 15 & valley silicon valley silicon layoffs & 35 & 0.66 \\
14 & voting voters just rankedchoice & 15 & valley silicon valley silicon layoffs & 35 & 0.69 \\
15 & values conservative development national & 14 & bitcoin currency crypto cryptocurrency & 74 & 0.65 \\
16 & computers machines let smarter & 14 & artificial intelligence intelligence artificial humans & 85 & 0.77 \\
17 & google android tech just & 13 & googles google deepmind google deepmind & 23 & 0.73 \\
18 & bostrom values machines human & 12 & artificial intelligence intelligence artificial humans & 85 & 0.75 \\
19 & nasa pen bostrom computers & 10 & artificial intelligence intelligence artificial humans & 85 & 0.73 \\
\botrule
\end{tabular}
\end{table}

\section{Power Analysis}
With $n_{\text{sim}}=30$ simulation runs and $n_{\text{Voat}}=30$ matched Voat validation windows, a two-sample design at $\alpha=0.01$ (aligned with our 99\% $t$-interval reporting for the main benchmark) has 80\% power to detect standardized effect sizes (Cohen's $d$) of about 0.91 (and 90\% power at about $d\approx 1.02$). Table~\ref{tab:power-analysis} reports observed effect sizes and the corresponding power for key metrics.

\begin{table}[htbp]
\centering
\caption{Power analysis for simulation vs.\ Voat comparison ($n_{\text{sim}}=30$, $n_{\text{Voat}}=30$, $\alpha=0.01$). MDES at 80\% power: $d=0.91$.}
\label{tab:power-analysis}
\small
\begin{tabular}{lrrrrc}
\toprule
\textbf{Metric} & \textbf{Sim Mean (SD)} & \textbf{Voat Mean (SD)} & \textbf{Cohen's $d$} & \textbf{Power} & \textbf{Overlap} \\
\midrule
\multicolumn{6}{l}{\textit{Activity Metrics}} \\
Root posts (threads) & 593 (32.1) & 569 (146.1) & 0.23 & 0.05 & \checkmark \\
Comments & 904 (46.4) & 733 (262.8) & 0.90 & 0.79 & -- \\
Unique users & 610 (29.9) & 591 (170.4) & 0.16 & 0.02 & \checkmark \\
Daily active users & 37.3 (1.4) & 32.7 (9.8) & 0.66 & 0.47 & \checkmark \\
Avg thread length & 2.53 (0.08) & 2.25 (0.22) & 1.68 & $>$0.99 & -- \\
\midrule
\multicolumn{6}{l}{\textit{Network Metrics}} \\
Network density & 0.0102 (0.0008) & 0.0065 (0.0014) & 3.16 & $>$0.99 & -- \\
Avg degree & 2.72 (0.12) & 2.31 (0.18) & 2.78 & $>$0.99 & -- \\
Core \% of LCC & 19.62 (2.05) & 5.14 (1.94) & 7.26 & $>$0.99 & -- \\
\midrule
\multicolumn{6}{l}{\textit{Toxicity}} \\
Mean toxicity & 0.1429 (0.0159) & 0.1186 (0.0205) & 1.33 & 0.99 & -- \\
\botrule
\end{tabular}
\end{table}
Metrics with overlapping CIs should be interpreted cautiously as approximate alignment at the current replication depth. Conversely, metrics with non-overlapping CIs show moderate-to-large effects with high power, indicating systematic divergences. Run-to-run variability is low (coefficients of variation below 12\% across the reported metrics), supporting the stability of the simulation dynamics under fixed parameters.

\endgroup

\end{document}